\documentclass[twocolumn]{aastex63}
\usepackage{CJKutf8} 
\usepackage{amsmath}
\usepackage{comment}
\usepackage{physics}
\usepackage{verbatim}

\usepackage{soul}

\submitjournal{AJ}

\begin{document}

\title{Systematically Measuring Ultra-Diffuse Galaxies (SMUDGes). VI. Nuclear Star Clusters}
  
\correspondingauthor{Mika Lambert}
\email{mlambert43@arizona.edu}

\author[0000-0002-2527-8899]{Mika Lambert}
\affiliation{Steward Observatory and Department of Astronomy, University of Arizona, 933 N. Cherry Ave., Tucson, AZ 85721, USA}

\author[0000-0002-7013-4392]{Donghyeon J. Khim}
\affiliation{Steward Observatory and Department of Astronomy, University of Arizona, 933 N. Cherry Ave., Tucson, AZ 85721, USA}

\author[0000-0002-5177-727X]{Dennis Zaritsky}
\affiliation{Steward Observatory and Department of Astronomy, University of Arizona, 933 N. Cherry Ave., Tucson, AZ 85721, USA}

\author[0000-0001-7618-8212]{Richard Donnerstein}
\affiliation{Steward Observatory and Department of Astronomy, University of Arizona, 933 N. Cherry Ave., Tucson, AZ 85721, USA}

\begin{abstract}

We present our photometric search for potential nuclear star clusters (NSCs) in ultra-diffuse galaxies (UDGs) as an extension of the SMUDGes catalog. We identify 325 SMUDGes galaxies with NSCs and, from the 144 with existing distance estimates, identify 33 NSC hosts as UDGs ($\mu_{0,g}$ $\ge$ 24 mag arcsec$^{-2}$, $r_e \ge 1.5$ kpc). The SMUDGes with NSCs lie on the galaxy red sequence, satisfy the NSC-host galaxy stellar mass relationship, have a mean NSC stellar mass fraction of 0.02 but reach as high as 0.1, have NSCs that are displaced from the host center with a standard deviation of 0.10$r_e$, and weakly favor higher density environments. All of these properties are consistent with previous results from higher surface brightness galaxy samples, allowing for at most a relatively weak dependence of NSC behavior with host galaxy surface brightness.

\end{abstract}


\section{Introduction}
\label{sec:intro}

The origin of massive, compact stellar populations in galaxies, whether those are globular clusters (GCs), massive black holes, or nuclear star clusters \citep[NSCs;][]{caldwell,binggeli,caldwell87,bothun}, remains poorly understood. Some models envision their formation through violent, extreme episodes of star formation \citep[e.g.,][]{mihos94,bekki01,kravtsov,kruijssen,renaud}, but such episodes may seem somewhat less likely in the low surface brightness galaxies that are the focus here. Other models relate the different population classes to each other, such as those that posit that NSCs form from the infall and merger of GCs \citep[e.g.,][]{tremaine,gnedin,salcedo,modak} and those where central massive black holes form from the dynamical collapse of NSCs \citep[e.g.,][]{begelman,miller,Antonini_2015}. Again, processes that may be common and relevant in massive, high surface brightness galaxies, such as dynamical friction, perhaps play a diminished role in the low mass, low surface brightness galaxies. Commonality of features makes it attractive to link these populations into a coherent scenario \citep[e.g.,][]{wehner,rossa,ferr,Fahrion_2021}.

At least regarding NSCs, basic constraints on any of these scenarios include the rates at which NSCs are found, the relationship between the NSC and host stellar masses and stellar populations, any connection to the host galaxy morphology, and the alignment of the NSC and its host's dynamical center --- all specified as a function of the relevant host galaxy properties. One key such property might be the host's central surface brightness, which presumably reflects the degree to which dissipation has concentrated matter toward the galaxy center, where an NSC or a central black hole would reside.

The richness of some of these constraints is already evident, for example, in the NSC occupation fraction (the fraction of galaxies that host an NSC). The occupation fraction varies with galaxy mass, rising and then falling as one proceeds from lower to higher mass galaxies \citep{neumayer, hoyer}. This behavior is quite distinct from that of the number of GCs in a galaxy, which varies proportionally with the host galaxy mass \citep{blakeslee,burkert} even to low masses \citep{2020Forbes_GC, 2022Zaritsky_GC}. One might naively have expected a close correspondence between the rate at which GCs and NSCs are found if NSCs are indeed formed from merged GCs, but the observed difference may highlight how certain details of the formation physics, such as the amplitude of dynamical friction, relate to the host galaxy properties. One can test scenarios along these lines and attempt to reproduce the NSC occupation fractions \citep[e.g.,][]{lotz, CDMB2009}. 

The incidence rate of GCs and the NSC occupation fraction are less well determined as a function of host galaxy surface brightness, but there are indications of a dependence of the NSC occupation fraction \citep{binggeli00,Lim2018}. Of course, the incidence rate is only one of various properties to explore. One could imagine that the typical mass of an NSC varies with host mass. In fact, such a trend has been observed \citep[e.g.,][]{2003ApJ...582L..79B,ferr,turner,2013ApJ...763...76S,denBrok2014}, and we are tempted to ask whether there is an analogous relation with the host surface brightness. 

Here we begin our exploration of the SMUDGes \citep{smudges,smudges2,smudges3,smudges5} set of ultra-diffuse galaxy (UDG) candidates to explore the nature of NSCs in low surface brightness galaxies. The value of UDGs to this topic is that they include the most massive, low surface brightness galaxies known \citep[cf.][]{2015vanDokkum} and thus may help us disentangle the roles of mass vs. surface brightness in shaping NSC properties. The value of the SMUDGes sample is that it is large and spans all environments, enabling us to also explore the possibility of a dependence of NSC properties on environment. This work, in which we focus on the identification of NSCs in SMUDGes galaxies is followed closely by Khim et al. (2023, in prep.) in which we extend the analysis to galaxies of somewhat higher surface brightness, but similar stellar masses, provided by the Sloan Digital Sky Survey \citep[SDSS;][]{SDSS_Kollmeier} and images from the DESI Legacy Survey \citep{dey}. That work will enable us to place the SMUDGes galaxies in a wider context, without the additional challenge of comparing across studies with disparate image quality and analysis methodology.

Uniform, well-defined criteria are essential for comparisons of NSC properties across host mass, surface brightness, or environment. As we will show, the data characteristics and criteria used to identify NSCs lead to large variations in sample properties. NSCs are broadly defined to be dense and massive stellar agglomerations that reside in or near the centers of galaxies that are brighter than the extrapolated surface brightness profile of the inner region of that galaxy \citep{neumayer}. However, consistently-applied, quantitative criteria do not exist for any of these defining characteristics. The purity and completeness of samples therefore varies among studies. Such (potential) differences between studies raise questions about any comparisons one would wish to pursue between, for example, cluster \citep{Lim2018} and field samples \citep{Carlsten_2022} of galaxies with NSCs. In this particular example, both studies find occupation fractions $\sim$ 20\%, but possible systematic differences leave open the question of whether this agreement is physically meaningful or fortuitous. Other studies have investigated the prevalence of NSCs in both early and late-type galaxies
\citep{Cote2006, Georgiev2014} and in dwarf galaxies \citep{Carlsten_2022}, but again comparisons among them remain challenging. 

As with every study, this one too has its weaknesses and strengths. Among the weaknesses relative to existing NSC studies is that we are working with shallower, lower resolution images than the state-of-the-art \citep[e.g., {\sl Hubble Space Telescope} images have been used that have a $\sim$ 3 mag deeper point source magnitude than our images;][]{Lim2018}. Thus, we only probe the bright end of the NSC luminosity function for the majority of our hosts, potentially suffer greater contamination, and obtain more uncertain photometric parameters for the NSCs themselves. On the other hand, the strength of this, and our sister study (Khim et al. 2023, in prep), is that our galaxy sample spans mass, environment in low luminosity, low surface brightness galaxies with consistent classification, thereby simplifying comparisons. 
In \S \ref{sec:method}, we describe our methodology. In \S \ref{sec:results}, we discuss the results as follows: 
1) an NSC classification catalog for the entire SMUDGes sample; 
2) our constraints on the relative co-centricity of NSCs and their hosts;
3) NSC properties and their relationship to host galaxy properties and;
4) any relation to the host galaxy environment. We use a standard WMAP9 cosmology \citep{wmap9}, although the results are insensitive to different choices of cosmological parameters at the level of current uncertainties, and magnitudes are from SDSS/DESI and are thus on the AB system \citep{oke1,oke2}.

\section{Methodology}
\label{sec:method}
\subsection{The Data}
We begin with the 6,805 visually-confirmed UDG candidates in the SMUDGes catalog \citep{smudges5}. 
These candidates were selected to have a low central surface brightness in the $g$-band, $\mu_{0,g}$ $\ge$ 24 mag arcsec$^{-2}$, and a large effective radius on the sky, r$_e \ge$ 5.3 arcsec.
In addition, we now impose a stricter color criterion than in the original work to remove likely background interlopers ($\sim$ 0.2 mag redder than the red sequence) and candidates with unphysical blue colors (0 $< g-r <0.8$; which removes 225 galaxies from the sample) and an angular size criterion to remove nearby galaxies that are unlikely to be UDGs and which corresponds to half our extracted image size ($r_e > 26$ arcsec; which removes 38 additional galaxies). We retain 6,542 galaxies to analyze.
The angular size cuts help our model fitting (see \S \ref{subsec: prep_model_fit}) by eliminating images that are either too small relative to our resolution or too large relative to the extracted image. These cuts do not easily translate to criteria on physical size because our candidates span a range of distances.

Importantly, for a study of NSCs in low surface brightness galaxies, the estimate of the central surface brightness of the galaxy used to define the catalog was calculated using S\'ersic model fitting where high surface brightness objects, including any potential NSC, were masked \citep{smudges3,smudges5}. This aspect of the parent survey, and also whether a limit on the value of the S\'ersic $n$ value is imposed (SMUDGes imposes $n<2$ in the fitting), can affect the sample selection and care must be taken if comparing results drawn from different catalogs of low surface brightness galaxies.

For our photometric analysis, we extract 200$\times$200 pixel (52.4 $\times$ 52.4 arcsec) $r$-band images of each of the 6,542 candidates from the 9th data release (DR9) of the Legacy Survey \citep{dey}.
These cutouts provide a sufficiently large field of view to include adequate background coverage.

\subsection{Preparing for Model Fitting}
\label{subsec: prep_model_fit}

To explore the morphology of each UDG candidate further than what was done in the SMUDGes catalog papers, and to determine if there is evidence of an NSC, 
we use the photometric model fitting software package GALFIT \citep{Peng_2010}. We take three preparatory steps before fitting any model. 

First, to assess the nature of a possible unresolved source near the center of each UDG candidate, we need the point spread function (PSF) of each image.
NSCs in our images are unresolved because their half light radii are typically $<$ 10 pc \citep{boker04,turner,Georgiev2014}, which corresponds to an angular size $<$ 1 arcsec for distances $>$ 2 Mpc. 
For each UDG candidate, we adopt the PSF model provided in the Legacy Survey for the relevant image.

Second, we generate an image of the pixel-by-pixel uncertainties, a $\sigma$-image, that is used to assess the likelihoods of the models GALFIT produces. We calculate the $\sigma$-image using the inverse-variance image provided by the Legacy Survey that was calculated using the image stack contributing to each pixel.

Finally, we 
isolate each UDG candidate from any surrounding bright sources that could affect the modeling. 
This is a complicated, iterative process that we describe in more detail below when discussing our model fitting procedure.

\subsection{Selecting Among Models}
\label{sec:galfit}
To identify NSCs in UDG candidates, we first assess whether there is evidence for a concentration of light beyond what can be described by a S\'ersic model with index $n \le 2$. If there is, then we assess the nature of that excess component. Because our targets are extremely diffuse and faint, the GALFIT fitting results are often highly sensitive to the adopted starting parameters for the model fitting, as well as to the presence of a central component, such as a bulge or NSC. To mitigate the impact of these factors on our results, we adopt a two-stage approach that we describe below and repeat the fitting multiple times with different adopted starting parameter values in each of the two stages. 
In the fitting, we use a convolution box set to half the length of the image, a magnitude zero-point of 22.5, originating from the definition of nanomaggies, and a plate scale of 0.262 arcsec pix$^{-1}$ for the Legacy Survey. We also summarize the procedure in Table \ref{tab:fitting}.

Our goal for the first fitting stage is to obtain a best-fit single S\'ersic model for each candidate UDG that is free from the influence of nearby objects or other stellar components (e.g., NSC or bulge). The results from this fit guide the fitting of the more complex models in the second stage.

\begin{figure}[ht]
    \centering
    \includegraphics[scale=0.13]{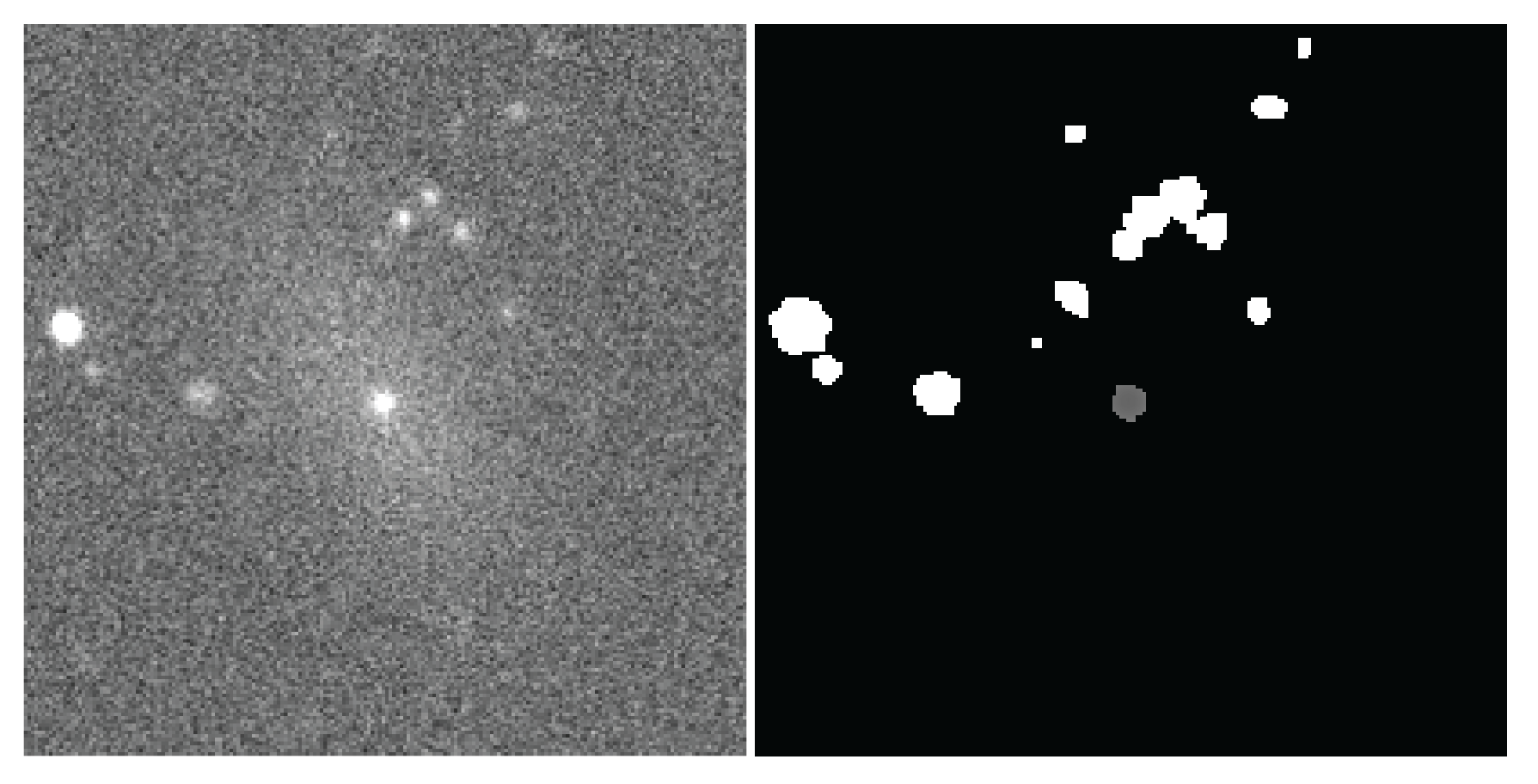}
    \caption{An example UDG candidate with a plausible NSC (SMDG0004118+163159, the second galaxy with what we will classify as an NSC in our right ascension ordered catalog) with several nearby contaminating sources is presented in the left panel. The mask in the right panel highlights most of the visible nearby sources with the central object shaded in gray indicating that it is unmasked for the final model fitting, but would masked in the initial model fitting (see text for details).}
    \label{fig:masks}
\end{figure}

We fit this single S\'ersic model, utilizing distinct image masks for each galaxy.
We create these masks (for an example, see Figure \ref{fig:masks}) using the Source Extractor Python library \citep[SEP, see][for details]{sep} based on \citep[SExtractor;][]{bertin}. We start by subtracting the spatially varying background generated by SEP.
We then identify objects defined as groupings of at least 5 adjacent pixels each with a flux that is 1.5$\sigma$ above the background. 
We mask these objects except for the galaxy itself. We remove from further consideration the four UDG candidates whose masked regions cover more than 50\% of the entire image area (leaving 6538 for study at this point) because these are poorly constrained in the fitting. Furthermore, in many cases, the fitting results would be strongly affected by an existing central NSC or bulge in the host galaxy. Thus, we augment the mask to include any central region of each candidate that contains at least 5 adjacent pixels that are 1.5 times (0.44 mag) brighter than $\mu =$ 24 mag arcsec$^{-2}$. This step introduces a potential bias against low luminosity NSCs that are not masked because they do not rise to this level and therefore are incorporated into the base S\'ersic model. We address sample completeness further below.

As we alluded to previously, to mitigate sensitivity to the adopted initial GALFIT fitting parameter values, we repeat the fitting six times using different values. We use combinations of two different effective radii (30 and 50 pix) and three surface brightness values at $r_e$ (25, 28, and 31 mag arcsec$^{-2}$). 
We calculate the reduced chi-squared statistic, $\chi^2_\nu$, within a circular region of radius 50 pixels ($\sim 13.1$ arcsec) centered on the image center and select the model with the smallest $\chi^2_\nu$ value.
The value of 50 pixels ensures that the bulk of the host galaxies falls within the evaluation region. We find that GALFIT occasionally produces a model fit with acceptable $\chi^2_\nu$ but unusually large final parameter uncertainties, even though other realizations (i.e. different initial parameters) result in fits with typical uncertainties. 
Because these solutions offer no meaningful constraints on the fit parameters, we only consider models where $r_e > 2\sigma_{r_e}$. We settled on this criterion after exploring a range of options, but these odd cases are clearly distinct from the remainder and various criteria could have been adopted without qualitatively affecting the results. In the rare case where all models for a particular galaxy fail this criterion (10 galaxies),
we consider the galaxy as having failed our fitting procedure. Additionally, we have 364 galaxies for which 
all models are rejected with at least 90\% confidence based on their $\chi_{\nu}^2$. 
These are also categorized as having failed our fitting procedure.

\begin{figure*}[ht]
    \centering
    \includegraphics[scale=0.47]{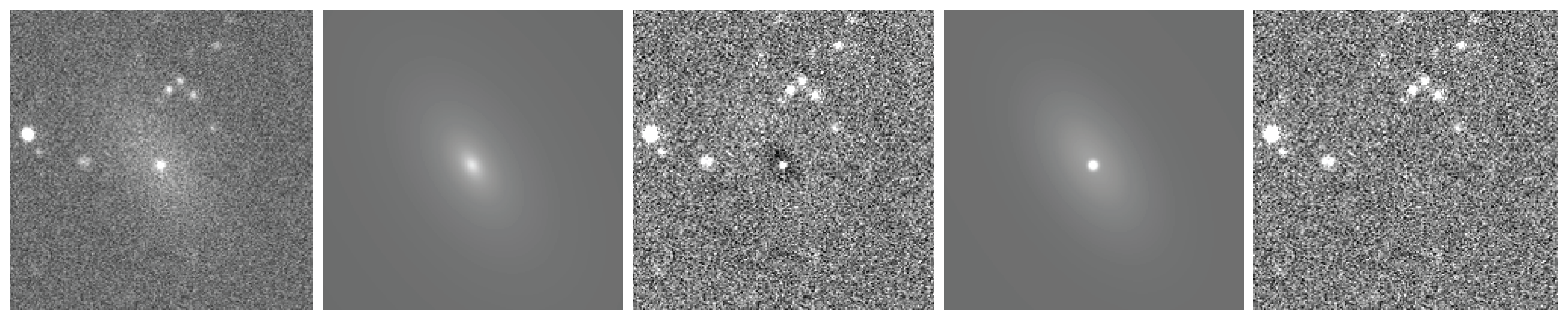}
    \caption{Model and residual images for SMDG0004118+163159. Left to right: the original image, single S\'ersic model, residual image using the single S\'ersic model, S\'ersic + PSF model, and residual image using the S\'ersic + PSF model. The need for a central unresolved source is evident when comparing the residual images.}
    \label{fig:models}
\end{figure*}

\begin{deluxetable}{lrr}
\caption{Model Summary}
\label{tab:fitting}
\tablehead{\colhead{1st component}&\colhead{2nd component}&\colhead{3rd component}}
\startdata
\multicolumn{3}{c}{\underbar{Initial GALFIT model (Stage 1)}} \\
\\
S$_1$ & ... & ... \\
\\
\multicolumn{3}{c}{\underbar{Final GALFIT model (Stage 2)}} \\
\\
S$_1$ & ... & ... \\
S$_1$ & S$_2$ & ... \\
S$_1$ & N$_{1,PSF}$ & ... \\
S$_1$ & S$_2$ & N$_{1,PSF}$\\
S$_1$ & N$_{1,PSF}$ & N$_{2,PSF}$ \\
\\
\enddata
\tablecomments{Columns indicate morphological components used in the six different model combinations. The initial parameters used for of the second stage are those of the best-fit model from the first stage.
We adopt two- or three-component models only when they are significantly favored over the one- or two-component models, respectively. Similarly, we only adopt models with unresolved sources when they are significantly favored over those with resolved components. A more comprehensive description can be found in Section \ref{sec:galfit}.
}
\end{deluxetable}

In the second stage, we unmask the center of the host galaxy and all other sources within $0.5r_e$ and compare new fits using five independent model classes for each candidate. These model classes consist of: 
1) a single S\'ersic profile, to model UDG candidates with no unresolved nuclear source; 
2) a pair of S\'ersic profiles, to model UDG candidates with an additional extended central source, such as a separate bulge component;
3) a S\'ersic profile plus a PSF profile, to model UDG candidates with an unresolved central source; 
4) a pair of S\'ersic profiles plus a PSF profile, to model UDG candidates with an extended central source, such as a separate bulge component, and an unresolved central source;
and
5) a S\'ersic profile plus two PSF profiles, to model UDG candidates with two unresolved central sources.
In the second and fourth model classes, we refer to the first S\'ersic component, intended to model the galaxy as a whole as S$_1$ and the second, more compact, component, which is intended to model any nuclear excess, as S$_2$.
For models with an unresolved central source, we refer to that component as N$_{1,PSF}$. In the fifth model class, we refer to the PSF component closer to the center of the S\'ersic component as N$_{1,PSF}$ and the further one as N$_{2,PSF}$.
Among the NSC targets, we have exclude those with a point source magnitude error in either $g$ or $r-$band larger than 0.2 mags.

As done in the first fitting stage, we fit each model class six times using different initial parameters. 
The initial parameters we adopt for S$_1$ are those from the best-fit model in the first stage, except for the central surface brightness.
For the initial brightness of S$_1$ and S$_2$ we take permutations of the best fit from the first stage and values that are three magnitudes fainter and brighter. 
For the size of S$_2$ we adopt a starting size of five pixels. We found no improvement in fitting when varying $r_e$ for S$_1$, so we simply use the value obtained in the first stage.

We provide GALFIT with the original UDG candidate image, the mask, the PSF, the $\sigma$-image, and a constraint file that sets the search range for each of the free parameters. The free parameters are the following: central positions, S\'ersic indices (n), effective radii ($r_e$), magnitudes (m), axis ratios (AR), and position angles (PA) for S$_1$ and S$_2$; and the position and amplitude for N$_{1,PSF}$ and N$_{2,PSF}$; and the background level. 
The constraint file provides initial parameter values and range limits for the chosen model parameters \citep{peng}. We constrain the centers of the various components to lie within a 40 by 40 pixel square centered on the candidate UDG. In some studies of NSCs, the centrality of the source is a critical criterion \citep[e.g.,][]{Cote2006,Georgiev2014}, but at this stage of our analysis we allow for significantly offset unresolved components. Because our sample consists of diffuse galaxies, we set the upper limit of the S\'ersic index to be 2.0 \citep[our UDG candidates have $\langle n \rangle < 1$; ][]{smudges3}. On the other hand, we allow the compact S\'ersic component (S$_2$) to have $n$ as large as 5.0 because such a component could be morphologically comparable to a bulge. We also set a lower limit on the axis ratio of the S\'ersic components of 0.3 to prevent unrealistically elongated models \citep[SMUDGes sample has an axis ratio threshold of $0.34 < b/a $; ][]{smudges5}.
Finally, we require that the effective radius of each S\'ersic component exceed 0.75 times the size of the PSF to ensure differentiation between what we consider to be a resolved component from an unresolved one.

Before comparing the resulting models, we exclude models that can be rejected with at least 90\% confidence given their $\chi^2_\nu$. 
Because we are primarily concerned with modeling the center of the UDG, the area within which we evaluate $\chi^2_\nu$ is now a circular region of radius $0.5r_e$, centered on the UDG candidate, where $r_e$ and the central position come from the single S\'ersic model fit from the first fitting stage.
Again, we only consider models where $r_e > 2\sigma_{r_e}$. If no models survive our $\chi^2_\nu$ and $r_e > 2\sigma_{r_e}$ criteria, then that galaxy is considered to have failed our fitting procedure.

Although $\chi^2_\nu$ values provide a goodness-of-fit measure, they are inappropriate for selecting among models of differing intrinsic complexity. Because a model with greater fitting freedom will naturally fit the data somewhat better, one must account for this additional flexibility when assessing whether there is statistical evidence in favor of the more complex model. The Akaike Information Criterion \citep[AIC;][]{AIC} is one formulation that incorporates a penalty for models with higher complexity.

We adopt a slight modification of the original AIC formulation that is referred to as the AICc criterion \citep{sugiura}, 
\begin{equation}
AICc= \chi^2+2p+ \frac{2p(p+1)}{N-p-1}
\end{equation}
where $p$ is the number of model parameters and $N$ is the number of data points that are fit,
to compare among models. 
This modification is appropriate when there are a small number of degrees of freedom and AICc will converge to the original AIC criterion as the degrees of freedom increase. The model with the smaller AICc value is the statistically preferred model although the greater the difference ($\Delta$AICc) the greater the confidence with which one can discriminate among them. Because AICc values describe likelihoods and are distributed like $\chi^2$, we are able to calculate the confidence level corresponding to any specific value of $\Delta$AICc. For our situation, a 2$\sigma$ confidence level corresponds to 
$\Delta$AICc
$= 11.83$. We adopt this threshold
to assess if the best-fitting two-component models are significantly preferred over the best-fitting one-component model. If it is, then we use this same threshold to assess if the model with S$_1$ + N$_{1,PSF}$ or S$_1$ + N$_{1,PSF}$ + N$_{2,PSF}$ is significantly preferred over that with S$_1$ + S$_2$. Figure \ref{fig:models} shows an example of a model and residual images created by GALFIT for the single S\'ersic model and the S$_1$ + N$_{1,PSF}$ model for SMDG0004118+163159. At this stage, we identify 
757 SMUDGes for which the S$_1 + $ N$_{1,PSF}$ or S$_1$ + N$_{1,PSF}$ + N$_{2,PSF}$ model is preferred over other competing models with greater than $2\sigma$ confidence.

\begin{figure}[ht]
    \centering
    \includegraphics[scale=0.47]{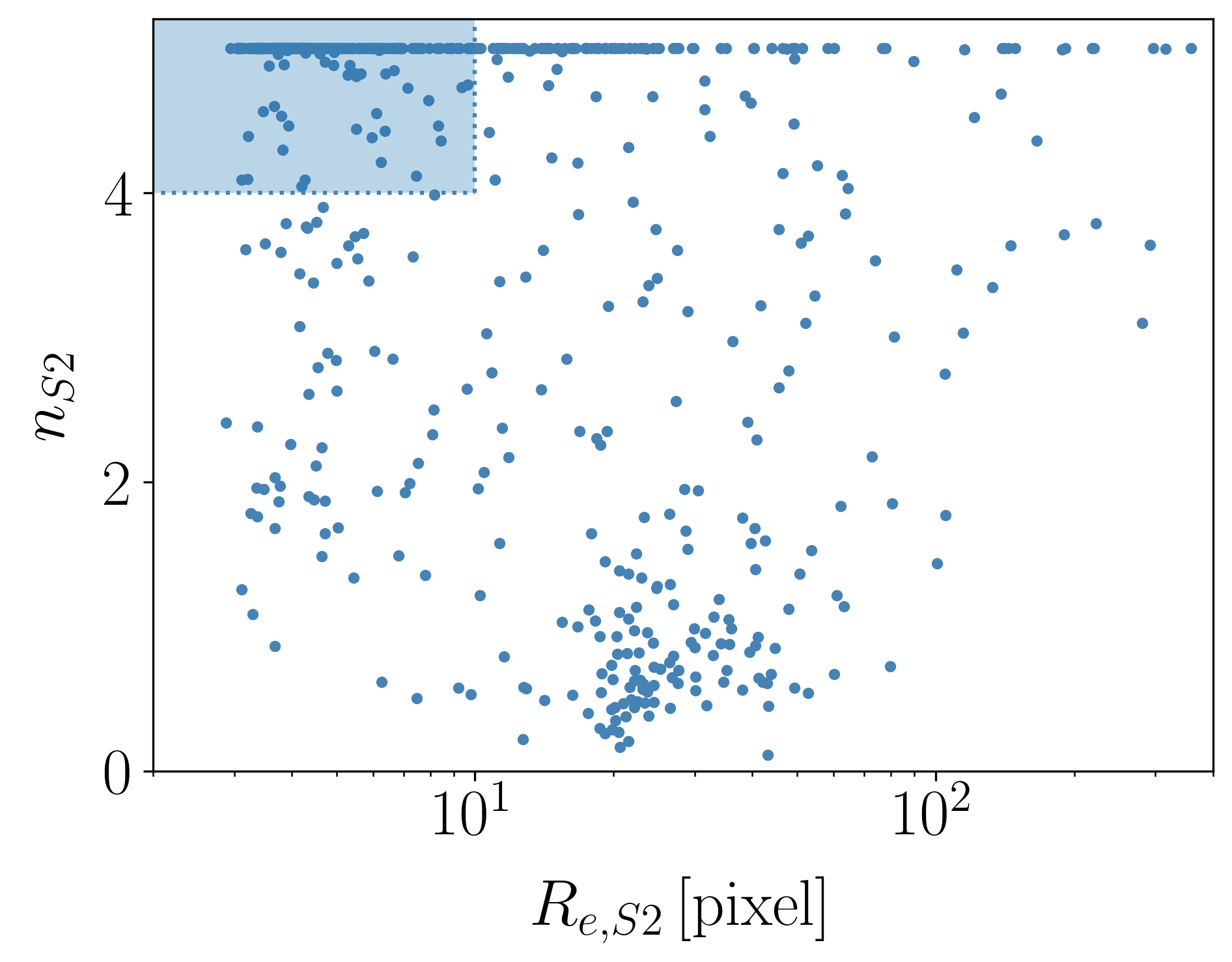}
    \caption{Distribution of $n$ and $r_e$ for S$_2$ components when two components are statistically favored over a single component, but no statistically significant difference exists between competing multi-component models. The box highlights the parameter region where, after visual examination, we reclassify systems as having unresolved sources.}
    \label{fig:s2_reclassify}
\end{figure}

Models where S$_2$ has a large $n$ value, $\ge 4$ and so greater than that corresponding to a de Vaucouleurs profile \citep{dev}, often appear visually indistinguishable from models with N$_{1,PSF}$. Furthermore, many of these S$_2$ components have small $r_e$, resulting in their observed morphology being dominated by the convolution with the PSF. For the vast majority of these, the models with N$_{1,PSF}$ are statistically favored over models with S$_2$, but not at the $2\sigma$ confidence level. After visual inspection, we decided to reclassify the subset of systems where two-component models are statistically favored over a one-component model at greater than 2$\sigma$ confidence and where that S$_2$ component has $n\ge4$, $r_e < 10$ pix (2.62 arcsec), and lies within 5 pixels of the competing N$_{1,PSF}$ as S$_1$ + N$_{1,PSF}$ (see Figure \ref{fig:s2_reclassify}). This increases the total number of SMUDGes within which we identify unresolved sources to 842. Nevertheless, the exact division, if one even exists, between a PSF-convolved unresolved source and a steep inner profile is not a simple issue \citep{cote}.

We present a summary of our classifications in Table \ref{tab:class} and an electronic version of a SMUDGes \citep{smudges5} line-matched catalog of our classifications as well as an example in Table \ref{tab:catalog}.
For each galaxy, we assign the most specific and complex model that is statistically preferred by the AICc criterion at the $>$ 2$\sigma$ confidence level, except for the S$_1$ classification which includes all cases where S$_1$ is favored, even if not at $>$ 2$\sigma$ confidence, and cases where another more complex model was statistically favored but not at beyond the $2\sigma$ confidence level. To provide more detail, in the case where we assign the S$_1 + $S$_2 + $N$_{1,PSF}$ classification it must be statistically preferred over both the S$_1 + $S$_2$ model and the S$_1 +$N$_{1,PSF}$ model at greater than the $2\sigma$ confidence level. In Table \ref{tab:class} we divide the classifications into objects with and without unresolved sources. The latter includes galaxies with indications of possible unresolved sources such as those where models with N$_{1,PSF}$ are preferred but not at the 2$\sigma$ level and those for which S$_1 + $S$_2 +$ N$_{1,PSF}$ is preferred at the $2\sigma$ level. We included the latter in this category because following visual inspection we decided that these are generally highly complex systems that are simply difficult to model and do not necessarily show evidence for NSCs. As a group these are identified as S$_1 +$ N$_{1,PSF}?$ in the Table.

\begin{deluxetable}{lrrrr}
\caption{Classification Summary}
\label{tab:class}
\tablehead{\colhead{Class}&\colhead{N}&\colhead{$r_p/r_e < 0.10$}&\colhead{N$_D$}&\colhead{$r_p/r_e < 0.10$}}
\startdata
\underbar{No unresolved}\\
\underbar{source}\\
\\
S$_1$ & 2979 & ... & 698 & ...\\
S$_1$+S$_2$ & 1558 & ...& 263& ...\\
S$_1$+N$_{1,PSF} ?$ & 795 & ... & 176& ...\\
Total& 5332& ... & 1137&...\\
\\
\underbar{Unresolved}\\
\underbar{source}\\
\\
S$_1$+N$_{1,PSF}$ & 673 & 242 & 221& 105\\
S$_1$+N$_{1,PSF}$ & 169 & 83& 63& 39\\
\ \ \ \ + N$_{2,PSF}$\\
Total&842&325&284&144\\
\\
\underbar{Failed fitting} & 364 & ... & 69& ...\\
\\
Total & 6538& 325 & 1490 & 144\\
\enddata
\tablecomments{Classifications for the SMUDGes candidates. The second column describes the classification outcomes for the SMUDGes that satisfy 0 $< g-r <0.8$ and $r_e < 26$ arcsec based on the original catalog \citep{smudges5} and are not heavily masked. The third column presents the numbers of those with identified unresolved sources that lie within a normalized projected separation from the S$_1$ component that is $< 0.10$, where the normalization is done using $r_e$ as measured from the initial fitting of the single S\'ersic model. The fourth column refers back to the second column and identifies the number of galaxies with estimated distances in the original catalog. The fifth column refers back to the objects in the fourth column and identifies the number of those that satisfy the projected separation criterion.}
\end{deluxetable}

\subsection{False Positives}

Among the 842 candidate UDGs for which the S$_1 + $ N$_{1,PSF}$ or S$_1$ + N$_{1,PSF}$ + N$_{2,PSF}$ models are preferred with $\ge 2\sigma$ confidence many show large offsets between $S_1$ and N$_{1,PSF}$ (for example, 44 have offsets larger than 20 pixels or 5.2 arcsec). This result raises the question of what constitutes a {\sl nuclear} star cluster.

\begin{figure}[ht]
    \centering
    \includegraphics[scale=0.355]{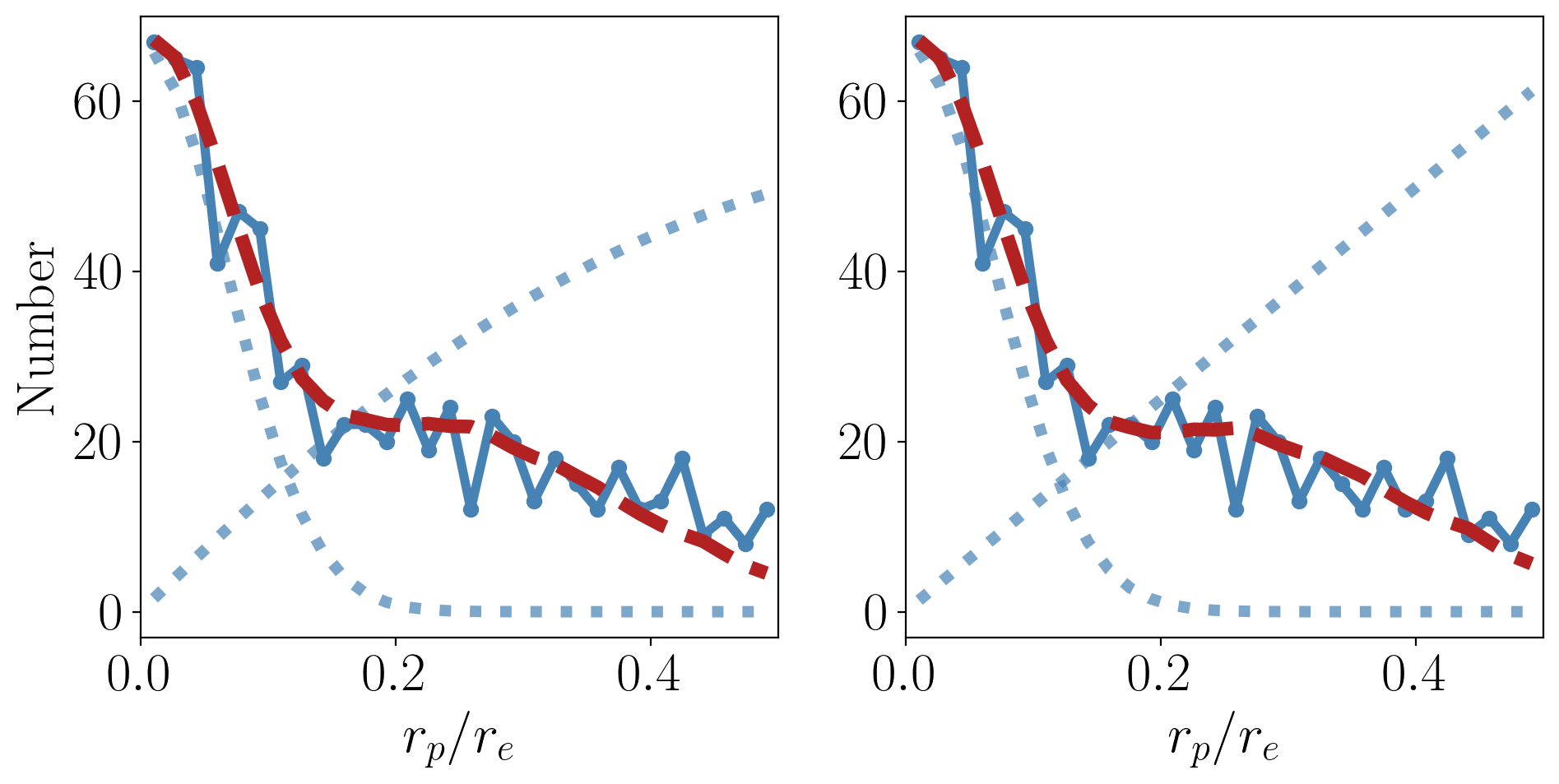}
    \caption{The distribution of normalized radial offsets between S$_1$ and N$_{1,PSF}$, $r_p/r_e$. Solid blue lines represent the data, the blue dotted lines show the models for the central Gaussian and background that together, modulated by the radial completeness, combine to produce the red dashed line. In the left panel, the background is assumed to follow the S$_1$ profile, while in the right panel, it is assumed to be randomly distributed on the sky. The models fit the data indistinguishably well, demonstrating that we cannot differentiate between the two background scenarios at the current time and that our conclusions are insensitive to this choice.}
    \label{fig:EMCEE_results}
\end{figure}

Our candidates span a range of projected offsets from the center of the host galaxy (Figure \ref{fig:EMCEE_results}). The offset distribution, plotted in normalized, distance independent terms of $r_p/r_e$, where $r_p$ is the projected separation between S$_1$ and N$_{1,PSF}$ and $r_e$ is the effective radius of the host galaxy (as measured in our initial fitting pass), appears to have two components: a concentrated central one and a far more extended one. The distribution is somewhat difficult to interpret because the positional offsets we allowed for GALFIT translate to different cutoffs in $r_p/r_e$ for each galaxy. Nevertheless, it suggests that our sample consists of populations we might call `true NSCs' and 'contamination'. The contaminating population may be a combination of sources physically associated with the galaxy, such as non-nuclear star clusters and star-forming regions, and physically unassociated sources, such as foreground stars and background galaxies. Critically, however, the details of how these populations are differentiated will impact certain questions related to the possibility of off-center nuclear star clusters, whether that be because they have stalled in their inward migration or been jostled off-center by a dynamical event.

To better understand the NSC and background populations, so that we can optimize our selection of NSCs, we explore a set of models for the radial distribution of all candidate NSCs in Figure \ref{fig:EMCEE_results}. In those models, we describe the NSC population alternatively as an empirically-motivated projected 1-D Gaussian distribution in normalized projected radial offsets, $r_p/r_e$, a 2-D Gaussian distribution in $r_p/r_e$ on the sky, or a 2-D exponential in $r_p/r_e$. The contaminating
population we describe alternatively as a S\'ersic distribution with the parameters of the host galaxy, assuming the contaminants come primarily from the galaxy itself, or a uniform distribution on the sky, assuming the contaminants are principally either foreground or background sources.
Finally, we also account for our radial completeness by evaluating the fraction of the sample at each radius for which our selection criteria would have allowed us to find an NSC. The radial completeness is affected both by our criteria that we only search for unresolved sources within $0.5r_e$ and allow GALFIT to explore positional offsets within a box of 40 by 40 pixels. 

We evaluate the parameters for each of these model combinations using 
a Bayesian approach and the Markov Chain Monte Carlo (MCMC) ensemble sampler called EMCEE \citep{EMCEE}. We model the distribution using different combinations of either a 1-D Gaussian or exponential primary distribution and a S\'ersic profile or uniform background secondary distribution. We adopt uniform priors for the amplitude and standard deviation of the central population and for the amplitude of the second component. We find a slight preference for the models that describe the distribution as a 1-D Gaussian central component plus S\'ersic-distributed contamination, but the exponential-distributed central component plus uniform background is nearly as good a fit (Figure \ref{fig:EMCEE_results}). As such, we conclude that we cannot distinguish whether the contaminating population is primarily within the host galaxy or unassociated on the basis of this data and fitting. This question will be reexamined in future work where we explore the extended population in greater detail. The contaminating population is potentially physically interesting because it might include clusters that are otherwise similar to NSCs but found at large radii.

\begin{figure*}[ht]
    \centering
    \includegraphics[scale=0.27]{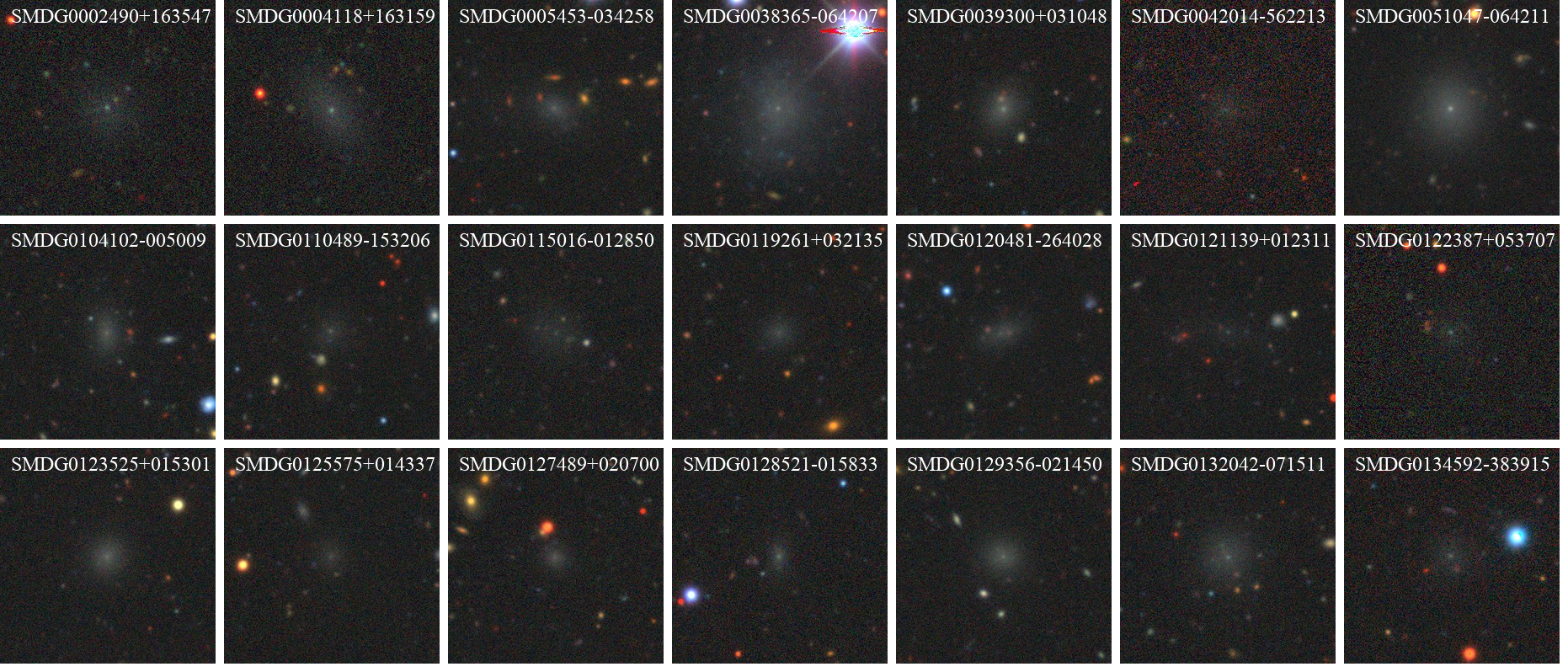}
    \caption{Mosaic of the first 21 SMUDGes sources in Right Ascension that have an identified NSC. Images span slightly over an arcmin on a side with North at the top and East to the left. These are drawn from the Legacy Surveys on-line viewer (\url{ https://www.legacysurvey.org/viewer}). Object labels are included at the top of each panel.}
    \label{fig:nscs}
\end{figure*}

To distinguish true NSCs from contamination, the choice of model becomes irrelevant because using either model we conclude that the contamination in our recovered NSC sample is 15\% (the percentage that we have set as our target) when we reject candidates with $r_p/r_e > 0.10$. Setting 0.10 as an upper limit on the normalized radial offset, we retain 325 UDG candidates with NSCs, and reject 517. Examples of our final NSC sample are presented in Figure \ref{fig:nscs} and this is the sample we present as NSC hosting SMUDGes. This severe selection demonstrates directly the impact of any imposed radial selection on the overall population. Consider that a slightly more permissive criterion of $r_p/r_e < 0.2$ results in $\sim$50\% more candidates (463), although a larger fraction of these will come from the contaminating population.

To highlight again how comparing among studies is fraught, we note that \cite{Poulain} accept the brightest unresolved source out to a radial offset of $0.5r_e$ as an NSC. After an initial reading of our results one would conclude that their sample is dominated by contaminants. However, their use of superior imaging data (both deeper and of higher angular resolution) might allow them to reject contamination far better than we can in our data. Without carefully examining both data sets and redoing parts of the analysis similarly, it is not possible to reach definitive conclusions regarding a comparison between our two studies.

Our strict radial selection cut ensures relatively high purity (defined to be 85\%) but excludes true NSCs that lie beyond this radial cut. We use our best-fit model for this component (a 1-D Gaussian with $\sigma = 0.10$) to calculate that 35\% of this component lies outside of $r_p/r_e = 0.10$. 
From the 325 candidate UDGs with NSCs at projected offsets less than $r_p/r_e = 0.10$, we calculate that 276 are true NSCs after correcting for the 15\% contamination and that this corresponds to a total population of UDG candidates with NSCs of 340 after correcting for systems that lie at $r_p/r_e > 0.10$. 

\subsection{False Negatives}

\begin{figure}[ht]
    \centering
    \includegraphics[scale=0.5]{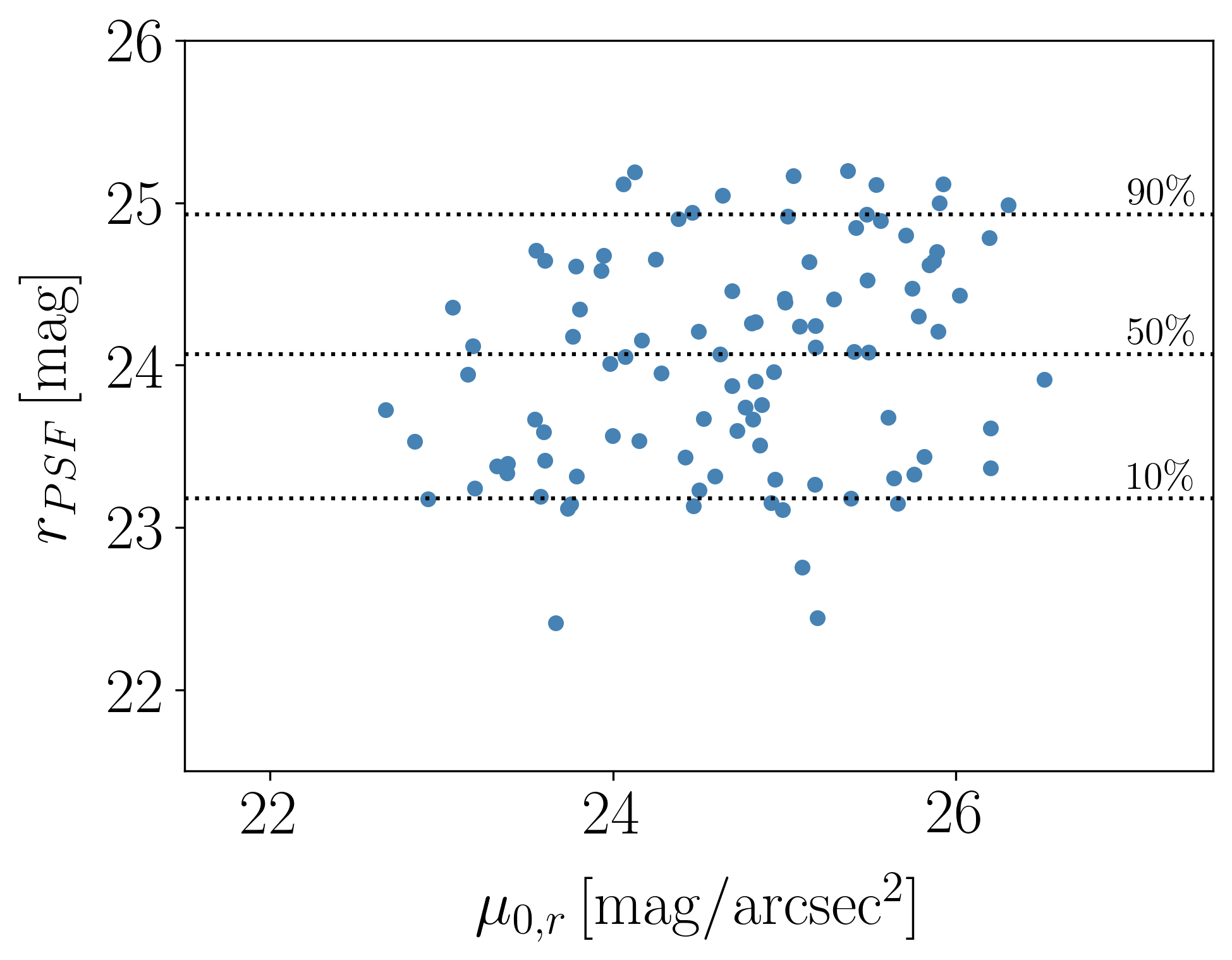}
    \caption{NSC detection limits in the $r$ band determined for 100 Single S\'ersic SMUDGes images as a function of $r-$band central surface brightness, $\mu_{0,r}$. Points represent the brightest simulated point source that was not recovered as an NSC by our procedure. Horizontal lines mark where we reach the corresponding incompleteness percentage. The three labeled values correspond to limiting magnitudes of $r_{PSF}$ of 23.2, 24.1, and 24.9 mag.}
    \label{fig:detection_sims}
\end{figure}

We classify as single component 2979 candidate UDGs.
Among this set, there may be some for which the NSCs fall below our detection limit and these are, as such, false negatives. To estimate our limiting magnitude,
we randomly select 100 galaxies that are preferred by a single S\'ersic model and insert artificial point sources of varying brightness ranging from 18 to 30 magnitudes, in intervals of 0.04 magnitudes, at the centers of the images.
We model these point sources using a Gaussian of 1.5 pixel width that extends out to 20 pixels and 
run our pipeline. When we do not recover the point source, there are two failure modes:
(1) the best-fit model is not the S\'ersic+PSF model; (2) 
the best-fit model is the S\'ersic+PSF model but has a confidence level of less than 2$\sigma$ ($\Delta$AICc $<$11.83). For each of the 100 galaxies, we set the magnitude of the brightest inserted point source that we fail to recover as the detection limit for that galaxy. In Figure \ref{fig:detection_sims} we present the set of these detection limits as a function of the host galaxy central surface brightness. The detection limits lie mostly between magnitudes of 23 to 25. We find that the limits correlate weakly with central surface brightness (confidence level 99.2\% and Spearman Rank correlation coefficient of 0.26). This result matches our intuition that it should be more difficult to detect an NSC in a galaxy that is itself intrinsically brighter in its center. However, given the large scatter in detection limits about this mean trend (Figure \ref{fig:detection_sims}), there must be other factors at play as well and we neglect the subtle, but real, surface brightness dependence in our subsequent qualitative discussion of completeness.

\begin{figure}[ht]
    \centering
    \includegraphics[scale=0.48]{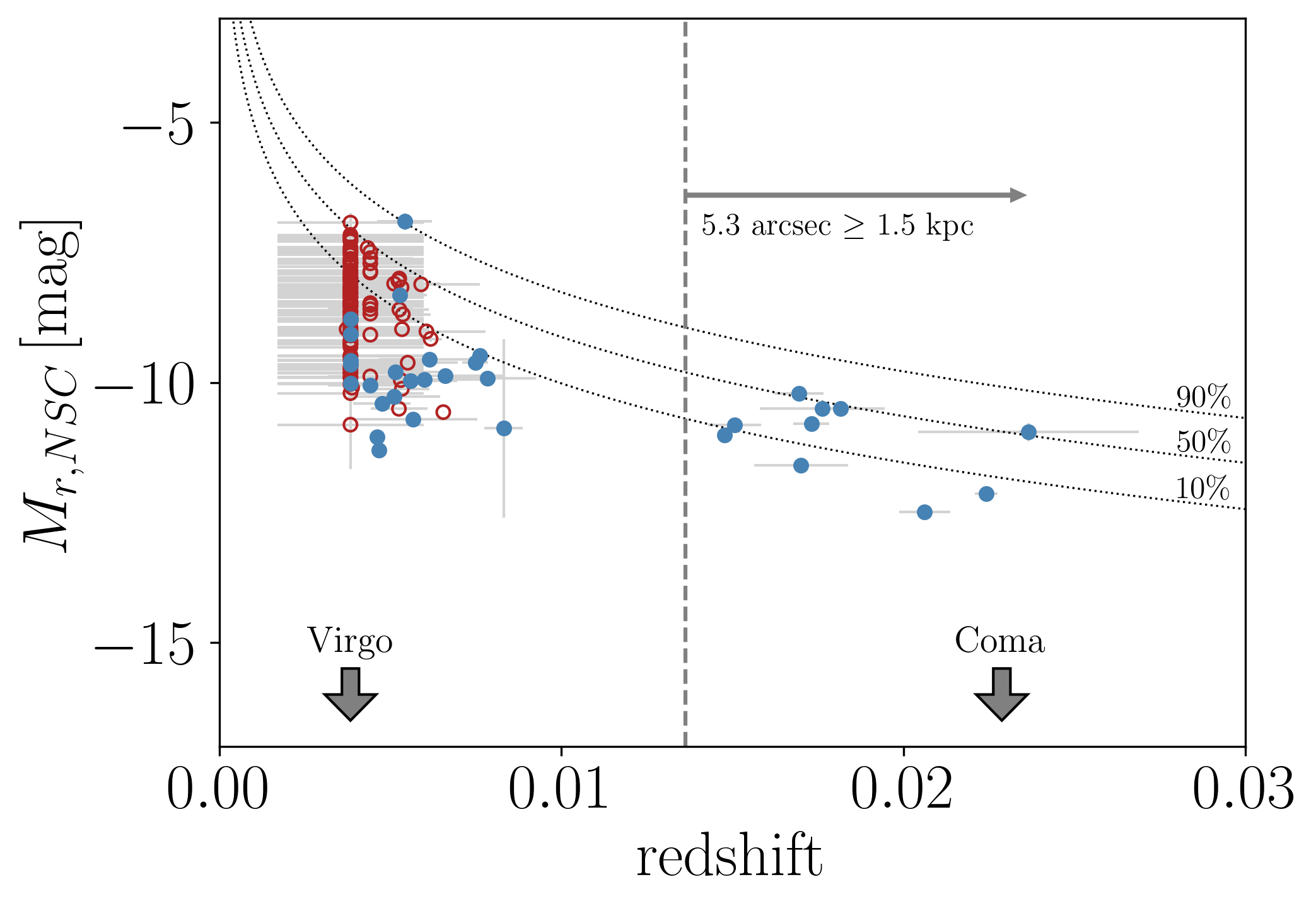}
    \caption{Absolute magnitude of NSCs, for SMUDGes with estimated distances, across redshift. Systems that are UDGs ($r_e \ge 1.5$ kpc; filled blue circles) and non-UDGs ($r_e < 1.5$ kpc; open red circles) are plotted. The curves designate completeness limits corresponding to the three incompleteness fractions shown in Figure \ref{fig:detection_sims}. We also include a vertical line that indicates the redshift beyond which any object in the SMUDGes catalog would satisfy the UDGs size criterion, and the redshifts corresponding to the Virgo and Coma clusters. }
    \label{fig:detection_limit}
\end{figure}

We now examine the consequences of using shallower, lower resolution data by comparing our classification results to those of \cite{Lim2018} for Coma galaxies using {\sl Hubble Space Telescope} images. 
\cite{Lim2018} presented 44 UDGs in their Table 1, 26 of which match SMUDGes. The majority of the unmatched galaxies are either of smaller angular extent than the SMUDGes limit or at a surface brightness where SMUDGes is incomplete. From the matched 26, \cite{Lim2018} concluded from visual inspection that five contain NSCs. Among these, we find an NSC in only one of those, but the normalized separation ($r_p/r_e =0.226$) is greater than our separation criterion ($r_p/r_e < 0.10$).
We attribute our failure to identify NSCs in the other four as the result of the difference between our magnitude limit of $\sim$24 mag and theirs of $\sim$ 27.4 mag \citep{Lim2018}. \cite{Lim2018} do not present photometry for their NSCs, so we cannot confirm that these are indeed fainter than our detection limit, but as we show in Figure \ref{fig:detection_limit}, our detection limit is likely to exclude nearly all NSCs by the time we are considering galaxies at the distance of the Coma cluster. This incompleteness also helps to explain why our occupation fraction (the fraction of candidates that we identify to host NSCs), which is globally $\sim 0.05$ (340/6538) for our sample, is so much lower than the $\gtrsim$ 0.20 that is commonly found
\citep[e.g.,][]{denBrok2014,Lim2018, 
Eigenthaler2018,Carlsten_2022}, although it may also reflect that our sample is not as constrained to high density environments and the occupation fraction is measured to be higher in high density environments \citep{Lim2018,sanchez,Poulain,Carlsten_2022}.

\begin{deluxetable*}{lrrrrrrrrr}
\caption{SMUDGes NSC Extension Catalog}
\label{tab:catalog}
\tablehead{
\colhead{SMDG Designation}&
\colhead{m$_{UDG,r}$}&
\colhead{m$_{NSC,r}$}&
\colhead{m$_{NSC,r,err}$}&
\colhead{m$_{NSC,g}$}&
\colhead{m$_{NSC,g,err}$}&
\colhead{$(g-r)_{UDG}$}&
\colhead{$r_e$}&
\colhead{$r_p/r_e$} &
\colhead{Class}\\
\colhead{ }&
\colhead{(mag)}&
\colhead{(mag)}&
\colhead{(mag)}&
\colhead{(mag)}&
\colhead{(mag)}&
\colhead{(mag)}&
\colhead{(arcsec)}&
\colhead{}&
}
\startdata
SMDG0000017+325141  &20.53&$-$99.00&$-$99.00&$-$99.00&$-$99.00&0.27& 5.32&$-$99.000&1\\
SMDG0000202$-$402435&20.95&$-$99.00&$-$99.00&$-$99.00&$-$99.00&0.30& 6.66&$-$99.000&6\\
SMDG0000334+165424  &18.24&$-$99.00&$-$99.00&$-$99.00&$-$99.00&0.55& 8.56&$-$99.000&2\\
SMDG0000453+305356  &19.90&$-$99.00&$-$99.00&$-$99.00&$-$99.00&0.18& 5.50&$-$99.000&2\\
SMDG0000473$-$040432&23.93&$-$99.00&$-$99.00&$-$99.00&$-$99.00&0.72&12.90&$-$99.000&7\\
\enddata
\tablecomments{We present here the first five lines of the full catalog, which is available in electronic form. Magnitudes refer to the S$_1$ and N$_{1,PSF}$ components of the best fit model, while colors and effective radii refer to those derived from the initial single S\'ersic fit, which are found to be more stable. Entries of $-99.00$ signal invalid values corresponding to SMUDGes galaxies for which an unresolved source is not identified. We retain these SMUDGes to maintain line-by-line matching with the \cite{smudges5} catalog. The numerical values in the Class column correspond, in order, to the classification categories in Table \ref{tab:class} ( S$_1 \equiv 1$, S$_1+$S$_2 \equiv 2$, S$_1 + $N$_{1,PSF}? \equiv 3$, S$_1+$N$_{1,PSF}\equiv 4$, S$_1+$S$_2+$N$_{1,PSF}\equiv 5$, failed fitting$\equiv 6$) and galaxies that were excluded prior to fitting ($\equiv 7$). Note that some S$_1+$N$_{1,PSF}$ and S$_1+$S$_2+$N$_{1,PSF}$ targets have photometric errors larger than 0.2 mags. These were classified as S$_1 \equiv 1$ in Table \ref{tab:class} and subsequently excluded from our analysis.}
\end{deluxetable*}

\section{Results}
\label{sec:results}

A principal product of this study is an NSC-related classification for each galaxy in the SMUDGes catalog and a measurement of the NSC properties when there is one. We present in 
Table \ref{tab:catalog} the first five lines of the full catalog, available electronically, where the objects are row matched with the \cite{smudges5} catalog. We provide a summary of the classifications in
Table \ref{tab:class}
by listing the number of objects in each of our classification classes, the number of which have an inferred unresolved source that meets our $r_p/r_e < 0.10$ criterion, the number in each class with distance estimates from the original SMUDGes catalog \citep{smudges5}, and the number of those with distance estimates that have an inferred unresolved source component that meets our $r_p/r_e$ criterion. 

\subsection{NSC Positional Offsets}

We have defined NSCs as the centrally located sub-population of unresolved sources coincident with our UDG candidates. 
The degree to which NSCs are truly found at the dynamical center of their host galaxy is somewhat difficult to address because NSCs are often required, by definition, to be at the galaxy's center \citep[e.g.,][]{2002AJ....123.1389B, Cote2006, 2011MNRAS.413.1875N}. For specific examples, we cite \cite{cote} which sets an offset upper limit of 0.02$r_e$ (about 20 pc for the typical distance in their sample) and \cite{neumayer} who propose an offset limit of 50 pc. In contrast, \cite{Poulain}, who allow for larger offsets, find NSCs out to 0.58$r_e$. 
\cite{binggeli00} identify some with even larger offsets. Offsets are potentially interesting to measure because they may be caused by dynamical interactions \citep{bellovary} and could help us measure the shape of the gravitational potential \citep{miller92,taga}.

Although we too have imposed such a requirement, based on the distribution of $r_p/r_e$, we do measure differences in the degree of alignment between S$_1$ and N$_{1,PSF}$. Careful examination of Figure \ref{fig:EMCEE_results} shows that the observed $r_p/r_e$ distribution does not peak at zero separation and we have measured a dispersion of 0.10.
To assess whether these findings reflect physical scatter in the co-centricity of the NSC and host or are simply the results of measurement errors always leading to positive offsets, we compare the measured distribution to the expected scatter arising simply from our observational uncertainties. When we adopt the GALFIT positional uncertainties in S$_1$ and N$_{1,PSF}$ for the individual observed systems and use those to randomly generate S$_1$ and N$_{1,PSF}$ pairs, we find a distribution of $r_p/r_e$ that also does not peak at zero but which has a dispersion of only 
0.02. From this result, we conclude that the observed scatter does constitute evidence of physical offsets. Our measured dispersion is in excellent agreement with the median offset measured in the MATLAS dwarf galaxy sample of 0.10$r_e$ \citep{Poulain}. Because the \cite{Poulain} sample has galaxies with a somewhat brighter central surface brightness than SMUDGes\footnote{The MATLAS sample \citep{Poulain} has a mean central surface in the $g-$band of 24.0 mag arcsec$^{-2}$, whereas the SMUDGes sample is defined to have a central surface brightness $\ge$ 24 mag arcsec$^{-2}$. Nevertheless, the MATLAS sample contains 53 satellites that qualify as UDGs and is an interesting complementary sample to ours.}, this agreement may be in conflict with findings of increasing offsets in lower surface brightness hosts \citep{binggeli00,Barazza}, although, as noted by \cite{Poulain} this measurement is complicated by the likely larger uncertainties in the measurements of the centers of lower surface brightness galaxies.

A different complication is that it is not clear which component is the better tracer of the dynamical center. It is possible that the photometric center of S$_1$ is precisely measured but that it does not reflect the dynamical center. All we can conclude is that the photometric center of S$_1$ and the position of N$_{1,PSF}$ show greater scatter than accounted for by the observational centroiding uncertainties. In fact, the entire dark matter halo may be offset leading to strong lopsidedness in the center of the galaxy \citep{prasad}.

\subsection{Nuclear and Non-nuclear Stellar Clusters: Host Galaxy Colors}

A related question is whether the unresolved sources associated with the more extended population include a substantial number of stellar clusters that are otherwise indistinguishable from those in the NSC population. Such a population would have important repercussions on models of NSC formation. 

To work with a set of unresolved sources that are mostly independent of what we have identified as the NSC population, we define a non-nuclear class in an analogous way as we did for the nuclear class, again focusing on purity. We use the model of the $r_p/r_e$ distribution but this time search for a lower limit on $r_p/r_e$ that ensures $\le 15$\% contamination of the non-nuclear population by the nuclear population. By setting that lower limit to be 0.38, we find 134 UDG candidates with unresolved sources that we define to be a `clean' sample of non-nuclear unresolved sources.

\begin{figure}[ht]
    \centering
    \includegraphics[scale=0.45]{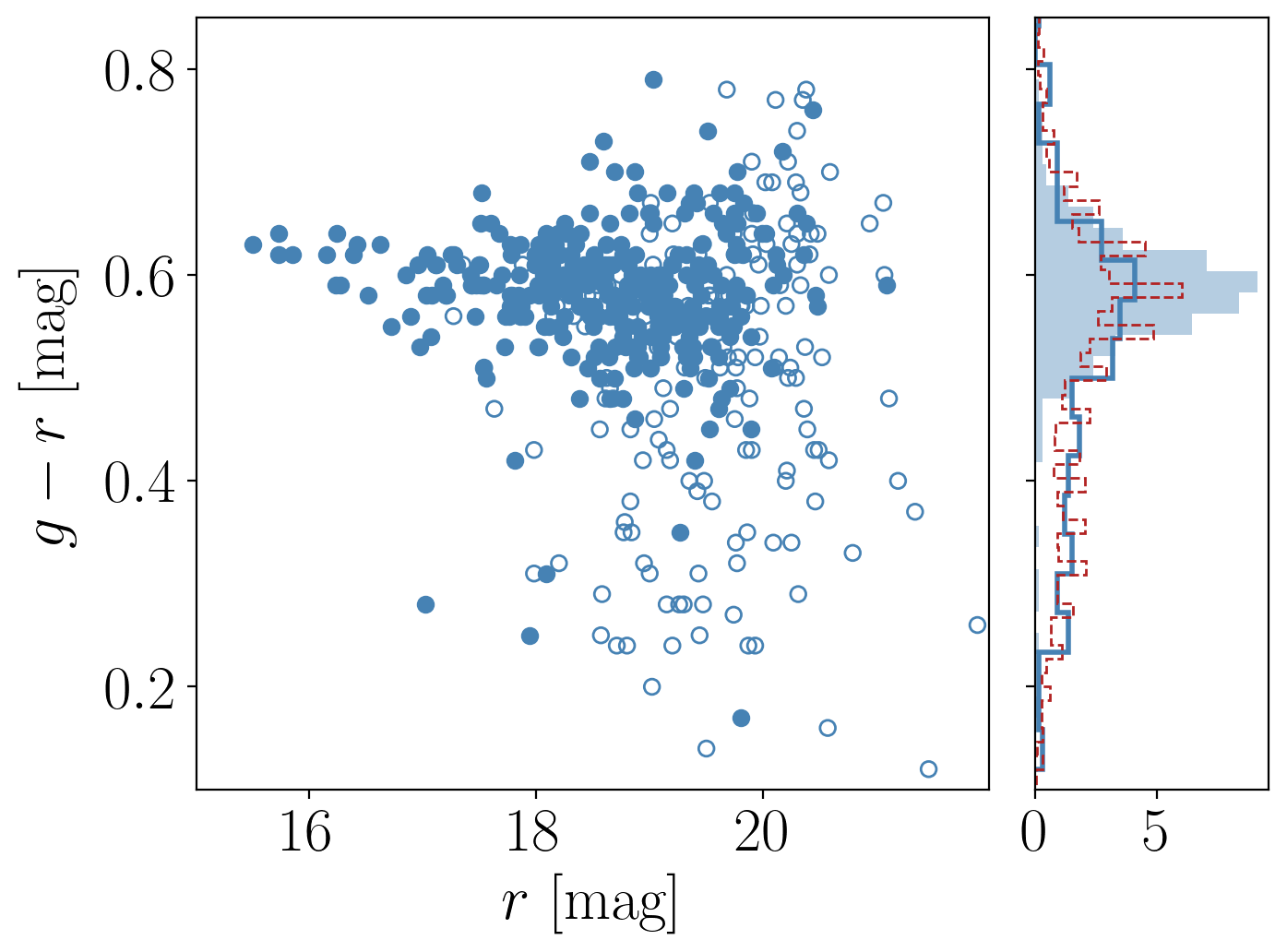}
    \caption{The distribution of host galaxy colors, $(g-r)_{UDG}$ from Table \ref{tab:catalog}, and apparent magnitudes for systems with unresolved sources with $r_p/r_e < 0.10$ (NSCs; filled circles) and $r_p/r_e > 0.38$ (non-NSCs; open circles). The right panel shows the normalized distributions in color of these two populations (filled histogram represents NSCs; unfilled blue, solid line represents non-NSCs), as well as that for the full SMUDGes sample (represented by the dashed red line).}
    \label{fig:color_mag}
\end{figure}

We find that the distributions in color differ markedly
(Figure \ref{fig:color_mag}).
NSC hosts have $g-r \sim$ 0.6, with a modest dispersion $\sim 0.05$, indicating that these hosts are predominantly on the red sequence. 
The number of NSC hosts that are much bluer ($g-r < 0.5$) is only 17 (out of the total of 325) and is smaller than the anticipated level of contamination (i.e. 15\% or 30 galaxies), suggesting that NSCs, as we have defined them, might only be found in red sequence hosts galaxies in our sample.
In contrast, the hosts of the non-nuclear unresolved-source population are distributed broadly in color. This flatter color distribution, and its similarity to the color distribution of the entire SMUDGes sample, further suggests that a significant fraction of this population is indeed contamination and not physically associated with the hosts.

The relative deficit of NSCs in our blue hosts raises the question of whether NSCs are exclusively a red galaxy phenomenon or reflect a selection bias that is working against us. Previous studies of late-type galaxies have found large occupation fractions \citep{Georgiev2014}, albeit not in hosts of this low surface brightness. We identify various challenges in identifying NSCs in blue SMUDGes. The hosts are more likely to be irregular and difficult to model, leading both to failures in the fitting and to greater statistical noise in the centroiding. The latter can both complicate our procedure for assessing the significance of an additional unresolved source and corrupt our measurement of the offset. Nevertheless, SMUDGes with color $0.4 < g-r < 0.5$ are not highly irregular and yet these already host relatively few NSCs, as can be seen by comparing the various distributions presented in Figure \ref{fig:color_mag}.

\subsection{NSC properties}

We now focus on the internal properties of NSCs in SMUDGes. We have two distinct populations to discuss. First, we draw inferences from our sample of 325 NSCs for which we do not have distance estimates, to discuss distance-independent aspects of the population. Their hosts are all galaxies of low central surface brightness, although it is likely that many will not satisfy the UDG physical size criterion. Second, we select only those hosts for which we do have distance estimates \citep[see][for a discussion of distance estimation]{smudges3,smudges5}, and then either focus on the physical properties of the 144 such systems or select only the 33 of those that meet the UDG size criterion ($r_e > 1.5$ kpc).

\begin{figure}[ht]
    \centering
    \includegraphics[scale=0.5]{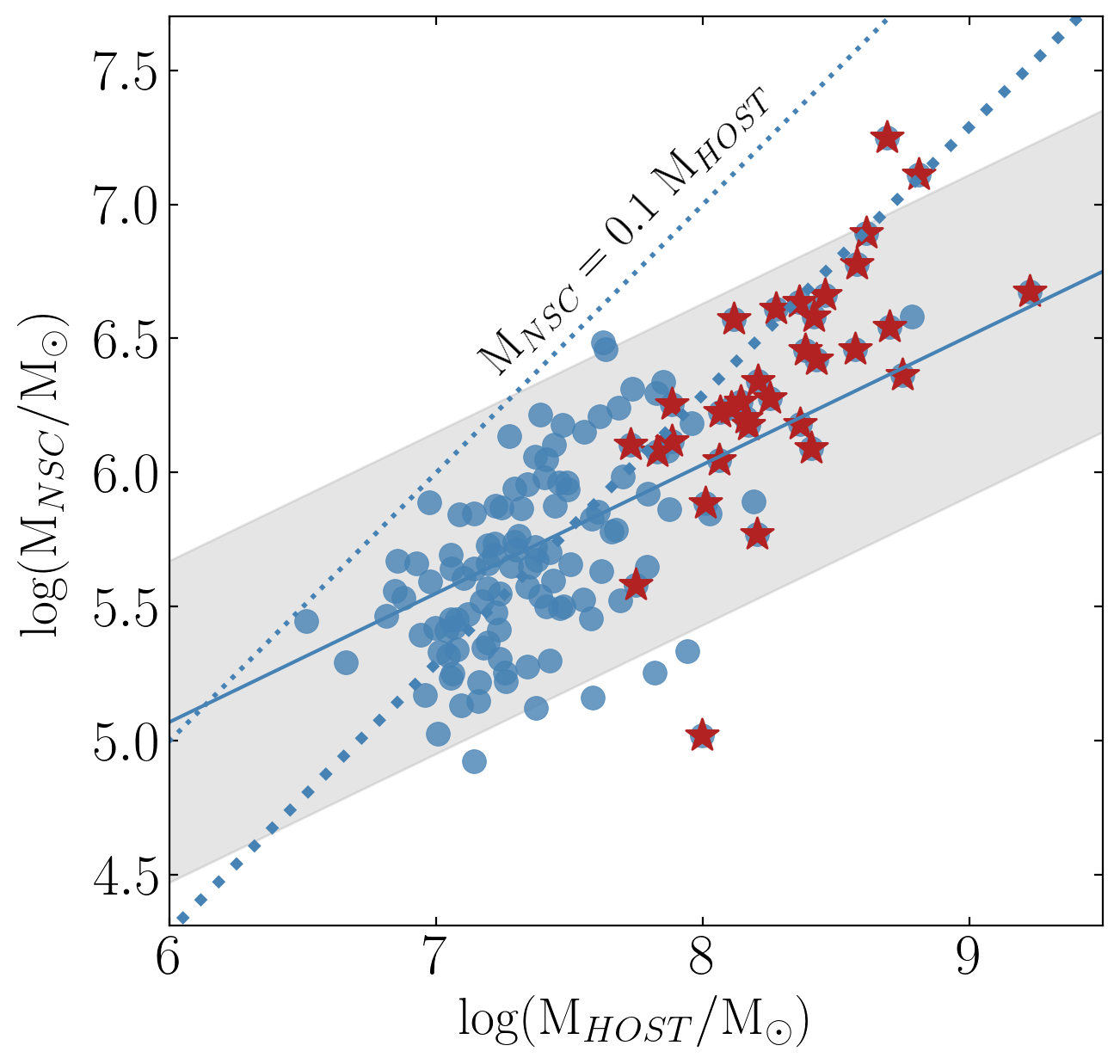}
    \caption{NSC stellar mass vs. host galaxy stellar mass. Stellar masses calculated assuming a fixed stellar mass-to-light ratio of 1.8 $M_\odot/L_\odot$. Solid line and shaded region represent the relationship and its uncertainty presented by \cite{neumayer}. The heavy dotted line represents a normalized 1:1 relation where M$_{NSC} = 0.019 M_{HOST}$. The thinner dotted line corresponds to M$_{NSC} = 0.1 M_{HOST}$ and represents the upper limit on the NSC mass fraction. Red stars represent UDGs. Blue dots represent the galaxies that do not meet the UDG physical size criterion.}
    \label{fig:mnsc-mstar}
\end{figure}

The luminosity (or corresponding stellar mass for similar mass-to-light ratios) of an NSC scales with that of the host galaxy \citep{neumayer}. 
We find this broadly holds for our sample (Figure \ref{fig:mnsc-mstar}). We adopt a stellar M/L of 1.8 M$_\odot$/L$_\odot$ for our transformation to stellar masses. 
The masses we derive, typically between $10^5$ and $10^{7}$ M$_\odot$ are within the previously measured range for NSCs \citep{neumayer}, although on the lower end as expected given that the stellar masses of our host sample is also lower than average. The data are mostly consistent with the previously published relationship \citep{neumayer}, although a steeper proportionality relation appears to fit the data better. However, recall that we are incomplete at lower NSC masses for more distant, hence typically larger and more massive hosts, which could help fill in the distribution in the lower right portion of the diagram. Furthermore, distance errors, and we expect that a significant fraction of the estimated redshifts ($\sim$ 30 \%) are incorrect \citep{smudges5}, will tend to have a preference for scattering objects to larger distances and proportionally higher masses along both axes. We conclude that our findings are consistent with the previous relation obtained with higher surface brightness galaxies. As such, we conclude that we find no systemic difference in the relationship between NSC and host galaxy stellar masses for low surface brightness galaxies relative to what was previously found for the more general NSC host population.

We find that the stellar mass fraction in NSCs can reach close to 0.1 in the most extreme objects, and is typically about 1/50th (Figure \ref{fig:mnsc-mstar}). This measurement is distance independent. At the upper end of the mass fractions, our findings are consistent with those of \cite{binggeli00} for Virgo dwarfs and mostly consistent with those of \cite{Poulain} for the MATLAS dwarfs. The latter do find objects with mass fractions $>$0.1, with one reaching 0.45, although they find only 6 in their sample of 508 with mass fraction $>$ 0.2. For the lower mass fractions our sample is incomplete, so we do not discuss that side of the distribution.
Our typical value of 0.02 is in excellent agreement with their median value of 0.017, again suggesting no strong dependence on host surface brightness.

\begin{figure*}[ht]
    \centering
    \includegraphics[scale=0.6]{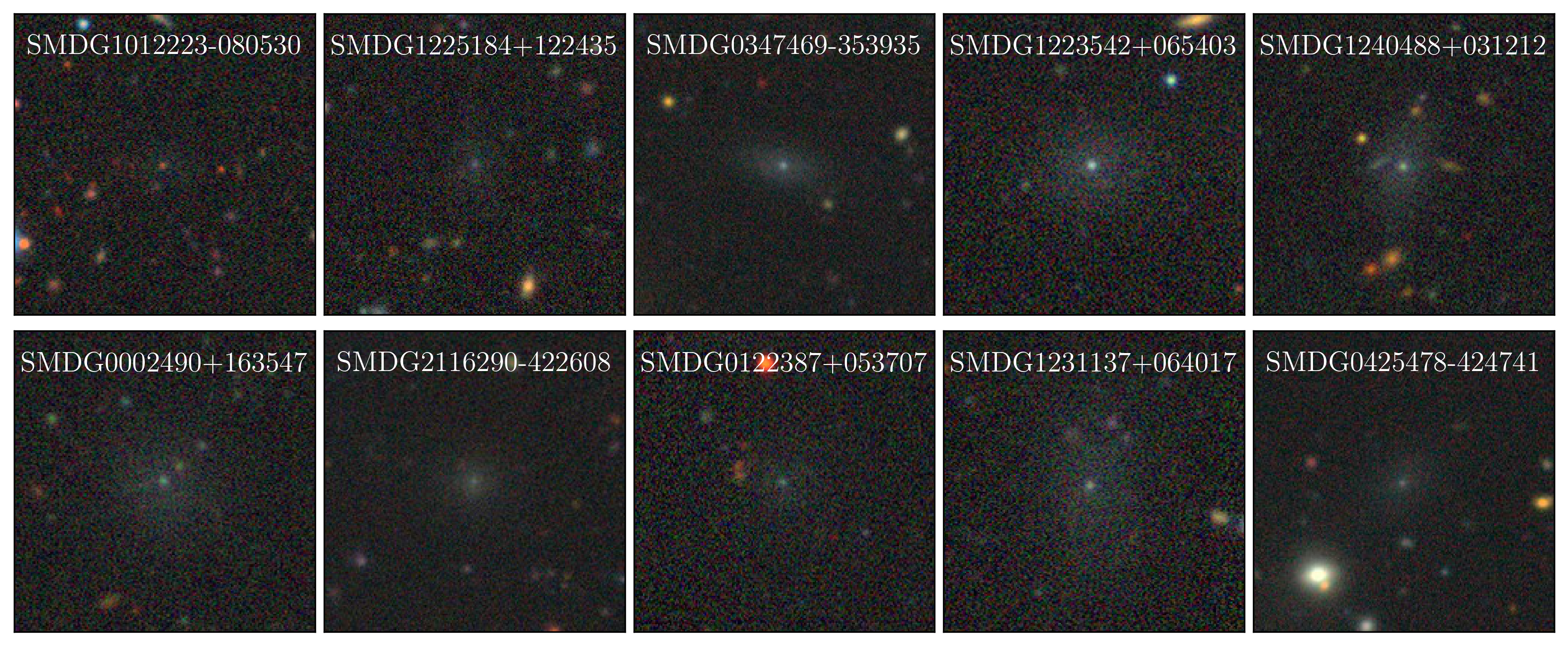}
    \caption{Mosaic of galaxies with extreme NSC mass fractions. The top row presents the five galaxies in our sample with the largest NSC mass fraction. The bottom row presents the five UDGs ($r_e \ge 1.5$ kpc) with the largest NSC mass fraction. 
    Images span slightly over an arcmin on a side with North at the top and East to the left. These are drawn from the Legacy Surveys on-line viewer. Object labels are included at the top of each panel.
    }
    \label{fig:extreme-massfrac}
\end{figure*}

The objects with the largest mass ratios appear to pose interesting constraints on formation models as it is difficult to envision how nearly 10\% of a galaxy's stellar mass could end up as an NSC either in the globular cluster infall model or in the central star formation model. 
In Figure \ref{fig:extreme-massfrac} we present the five galaxies and the five UDGs with the most extreme NSC mass fractions in our sample. 
In our entire sample, where we can use galaxies without estimated distances because we are considering only the ratio of stellar masses, we do find that the most extreme galaxies, such as those shown in the upper row of the Figure, have slightly more than 10\% of their stellar mass in their NSCs. For the most extreme UDGs that percentage drops to less than 3\%. Given the small number of confirmed UDGs in the sample, we do not yet know if this represents a real difference between UDGs and other galaxies or if we have simply not yet sampled the high fraction tail sufficiently well. In closing, we note that the NSC in one of the UDGs (SMDG0430339-052909) has a measured color that is anomalously red and so we suspect that this is an unresolved background galaxy rather than an NSC. Having one contaminant among the ten galaxies shown in the Figure is consistent with our anticipated 15\% contamination rate.

\subsection{Environment}

The hosts of NSCs in the SMUDGes sample are almost exclusively on the red sequence (Figure 
\ref{fig:color_mag}). Given the connection between UDG color and environment \citep{prole,kadowaki21}, this finding may point to an environmental dependence of the occupation fraction, in the sense as others have found \citep{Lim2018,sanchez,Poulain,Carlsten_2022}. 

\begin{figure}
    \centering
    \includegraphics[scale=0.4]{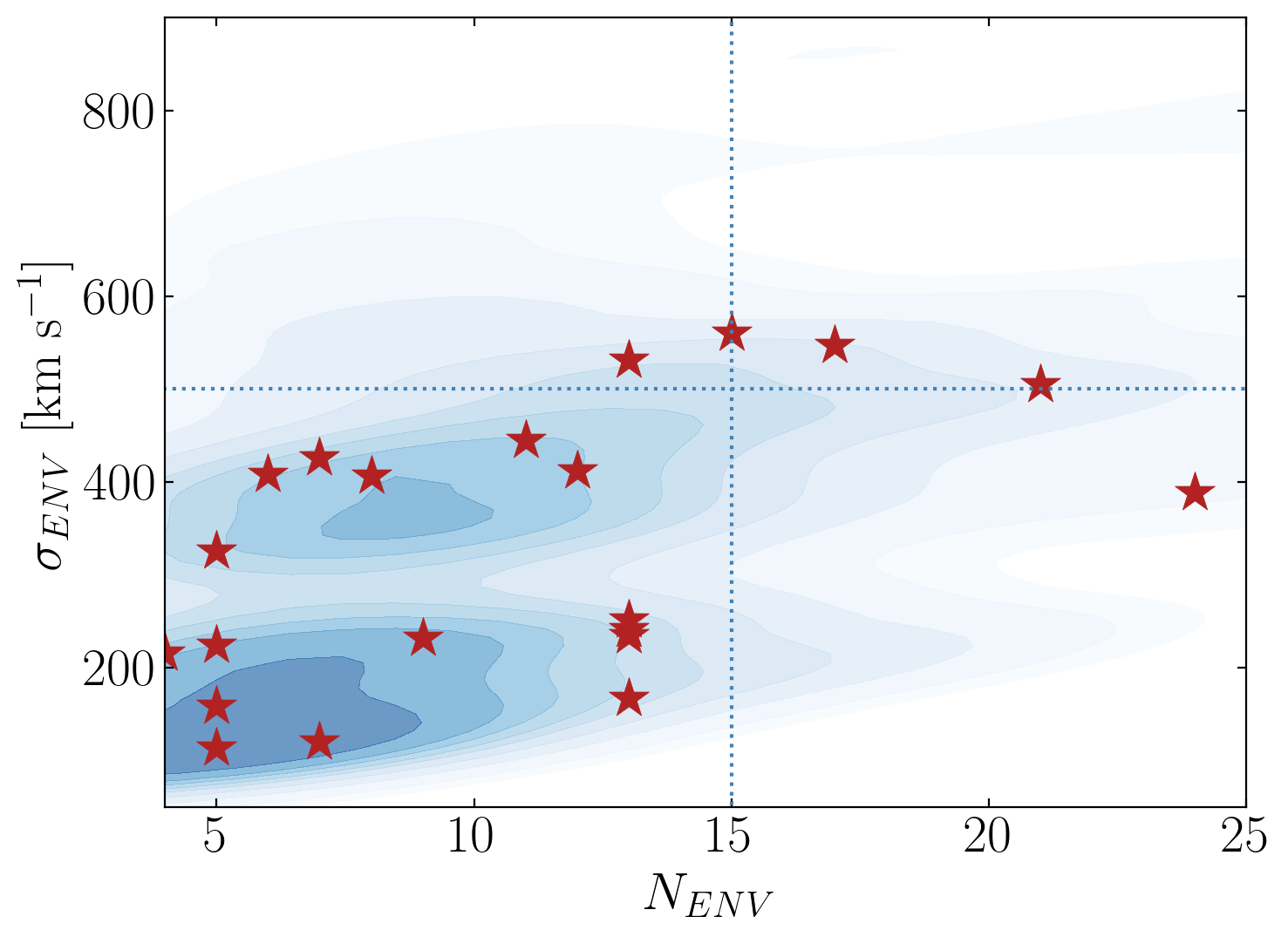}
   \caption{Environmental properties of UDGs with and without NSCs. We compare the local measures of environment, $\sigma_{ENV}$ and $N_{ENV}$, measures of the local velocity dispersion and richness provided by \cite{smudges5}, for UDGs with NSCs (red stars) and UDGs without NSCs (smoothed blue distribution). The dotted lines mark the regions set by \cite{smudges5} to differentiate low ($N_{ENV} < 15$ and $\sigma_{ENV}<500$) and high ($N_{ENV} > 15$ and $\sigma_{ENV}>500$) density environments.}
    \label{fig:environment}
\end{figure}

In Figure \ref{fig:environment} we compare the distribution of UDGs with NSCs to those without NSCs in the environmental parameter space defined by the estimated velocity dispersion, $\sigma_{ENV}$, and the richness, $N_{ENV}$, of the hosting overdensity that are provided by \cite{smudges5}. Those parameters are known to be highly uncertain measures of environment, so \cite{smudges5} suggest using them in combination to define poor and rich environments (the lower left and upper right quadrants in Figure \ref{fig:environment}, respectively). We are limited in this comparison by the small number of UDGs with NSCs (33), but the visual impression from the Figure is that UDGs with NSCs tend to higher values of $\sigma_{ENV}$ than the overall sample. Although the sense of this behavior is as previously observed \citep[e.g.,][]{Lim2018,sanchez,Poulain}, statistical tests (comparison of means, medians, KS test) do not yield a statistically significant detection of a difference in the two populations presented here. 

Confirming this preliminary result is important because it would extend the relation between NSC occupation fraction and environment to that of groups and the field and to UDGs. This will require spectroscopic redshifts of many more SMUDGes galaxies to provide distances and enable us to convert angular measurements to physical ones.

\section{Summary}
\label{sec:summary}
We present the results of our photometric search for potential nuclear star clusters (NSCs) hosted by 
the ultra-diffuse galaxy candidates in the SMUDGes catalog \citep{smudges5}. Using $r$-band images from DR9 of the Legacy Survey \citep{dey}, we develop an algorithm to statistically determine if additional photometric components beyond a single S\'ersic model within 0.5$r_e$ are needed, and then if among those components there is one that is best modeled as an unresolved object. We find that slightly over half of the SMUDGes sample does show evidence for additional components. Among those, we identify 1059 for which we find with 90\% confidence the need for an unresolved source. 

The distribution in projected radius of these unresolved sources shows a peak near zero separation and a second more extended component. We explore models and quantify the nature of the two components, attributing only the central component to NSCs. We use our models of the radial distribution to define a maximum projected separation for our defined NSC sample that ensures an 85\% pure sample of candidate NSCs (0.10$r_e$). 

We explore our incompleteness using simulations and establish 
that we are significantly incomplete due to our relatively shallow imaging (NSC magnitude limits ranging from $\sim 23$ to 25 mag depending on the specifics of the host galaxy and surroundings). Nevertheless, we are able to confirm with confidence 325 SMUDGes galaxies with NSCs, 144 of which also have estimated distances provided by \cite{smudges5}. Among those 144, we confirm 33 as UDGs with NSCs.

Despite our identification of NSCs as a population that is closely aligned with the center of their host galaxy, 
we find that the observed scatter of positional offsets between NSCs and their hosts is greater than expected from measurement errors alone. Our estimate of the dispersion in offsets (0.10$r_e$) is in good agreement with the median offset measured from MATLAS dwarf galaxies \citep{Poulain}. Such offsets could be used to constrain formation scenarios and models of the host's gravitational potential \citep{miller92,taga,bellovary}.

We find that our sample of NSCs is hosted almost exclusively by galaxies on the red sequence. The number of NSCs found in bluer hosts is consistent with the expected level of contamination. This result may reflect the color-environment trend identified for such galaxies \citep{prole,kadowaki21} and the greater NSC occupation fraction in denser environments \citep{Lim2018,sanchez,Poulain,Carlsten_2022}. Unresolved sources away from the nucleus are found in hosts whose color distribution matches that of the SMUDGes sources overall, suggesting no physical connection. We discuss some possible selection biases against our finding NSCs in blue hosts.

Despite our focus on low surface brightness galaxies, and UDGs in particular, we find that the NSCs
in our sample fall on the NSC-host galaxy stellar mass relationship found previously \citep{neumayer} from higher surface brightness objects. There is potentially a deviation for the largest objects in our sample (the UDGs) but our strong selection effects and the currently necessity of comparing across studies rather than within a single study limits our conclusions in this regard.
The typical NSC in our sample contains about 0.02 of the total stellar mass of the host galaxy, although the most extreme objects reach a fraction of nearly 0.1. These results are in agreement with previous studies of NSCs in somewhat different galaxy samples \citep{binggeli00,Poulain}.

Finally, we search for possible environmental effects in the NSC population. Despite this being principally a field sample, we find a suggestion that NSCs are more likely in UDGs in higher density environments, in agreement with previous results \citep{Lim2018,sanchez,Poulain,Carlsten_2022}. However, quantitative analysis of this trend in our data does not yet yield statistically significant results. Increasing the sample of UDGs with NSCs, by having spectroscopic redshifts of a larger fraction of the SMUDGes NSC sample, will allow us to further assess this possibility in the future.

The SMUDGes catalog and the NSC extension catalog provided here enable ongoing work on the nature of NSCs and their hosts, extending the latter to low surface brightness, physically large galaxies. This sample will benefit greatly from complementary high angular resolution imaging to come from surveys carried out with the {\sl Euclid} and {\sl Nancy Grace Roman} telescopes, spectroscopy from highly multiplexed surveys such as DESI, and from continued dedicated follow-up observations. 

\begin{acknowledgments}

The authors acknowledge financial support from NSF AST-1713841 and AST-2006785. An allocation of computer time from the UA Research Computing High Performance Computing (HPC) at the University of Arizona and the prompt assistance of the associated computer support group is gratefully acknowledged.
The authors also thank Min-Su Shin for their on-line description on scaling fits files to png images using their img\_scale.py module.

\end{acknowledgments}

\software{
Astropy              \citep{astropy1, astropy2},
GALFIT               \citep{peng},
Matplotlib           \citep{matplotlib},
NumPy                \citep{numpy},
pandas               \citep{pandas},
sep                  \citep{sep},
Source Extractor     \citep{bertin},
SciPy                \citep{scipy1, scipy2},
EMCEE                 \citep{FM2013}
skimage                \citep{scikit-image}
}

\bibliography{references.bib}{}

\begin{thebibliography}{}
\expandafter\ifx\csname natexlab\endcsname\relax\def\natexlab#1{#1}\fi
\providecommand{\url}[1]{\href{#1}{#1}}
\providecommand{\dodoi}[1]{doi:~\href{http://doi.org/#1}{\nolinkurl{#1}}}
\providecommand{\doeprint}[1]{\href{http://ascl.net/#1}{\nolinkurl{http://ascl.net/#1}}}
\providecommand{\doarXiv}[1]{\href{https://arxiv.org/abs/#1}{\nolinkurl{https://arxiv.org/abs/#1}}}

\bibitem[{Akaike(1974)}]{AIC}
Akaike, H. 1974, IEEE Transactions on Automatic Control, 19, 716,
  \dodoi{10.1109/TAC.1974.1100705}

\bibitem[{{Antonini} {et~al.}(2015){Antonini}, {Barausse}, \&
  {Silk}}]{Antonini_2015}
{Antonini}, F., {Barausse}, E., \& {Silk}, J. 2015, \apj, 812, 72,
  \dodoi{10.1088/0004-637X/812/1/72}

\bibitem[{{Astropy Collaboration} {et~al.}(2013){Astropy Collaboration},
  {Robitaille}, {Tollerud}, {Greenfield}, {Droettboom}, {Bray}, {Aldcroft},
  {Davis}, {Ginsburg}, {Price-Whelan}, {Kerzendorf}, {Conley}, {Crighton},
  {Barbary}, {Muna}, {Ferguson}, {Grollier}, {Parikh}, {Nair}, {Unther},
  {Deil}, {Woillez}, {Conseil}, {Kramer}, {Turner}, {Singer}, {Fox}, {Weaver},
  {Zabalza}, {Edwards}, {Azalee Bostroem}, {Burke}, {Casey}, {Crawford},
  {Dencheva}, {Ely}, {Jenness}, {Labrie}, {Lim}, {Pierfederici}, {Pontzen},
  {Ptak}, {Refsdal}, {Servillat}, \& {Streicher}}]{astropy1}
{Astropy Collaboration}, {Robitaille}, T.~P., {Tollerud}, E.~J., {et~al.} 2013,
  \aap, 558, A33, \dodoi{10.1051/0004-6361/201322068}

\bibitem[{{Astropy Collaboration} {et~al.}(2018){Astropy Collaboration},
  {Price-Whelan}, {Sip{\H{o}}cz}, {G{\"u}nther}, {Lim}, {Crawford}, {Conseil},
  {Shupe}, {Craig}, {Dencheva}, {Ginsburg}, {VanderPlas}, {Bradley},
  {P{\'e}rez-Su{\'a}rez}, {de Val-Borro}, {Aldcroft}, {Cruz}, {Robitaille},
  {Tollerud}, {Ardelean}, {Babej}, {Bach}, {Bachetti}, {Bakanov}, {Bamford},
  {Barentsen}, {Barmby}, {Baumbach}, {Berry}, {Biscani}, {Boquien}, {Bostroem},
  {Bouma}, {Brammer}, {Bray}, {Breytenbach}, {Buddelmeijer}, {Burke},
  {Calderone}, {Cano Rodr{\'\i}guez}, {Cara}, {Cardoso}, {Cheedella}, {Copin},
  {Corrales}, {Crichton}, {D'Avella}, {Deil}, {Depagne}, {Dietrich}, {Donath},
  {Droettboom}, {Earl}, {Erben}, {Fabbro}, {Ferreira}, {Finethy}, {Fox},
  {Garrison}, {Gibbons}, {Goldstein}, {Gommers}, {Greco}, {Greenfield},
  {Groener}, {Grollier}, {Hagen}, {Hirst}, {Homeier}, {Horton}, {Hosseinzadeh},
  {Hu}, {Hunkeler}, {Ivezi{\'c}}, {Jain}, {Jenness}, {Kanarek}, {Kendrew},
  {Kern}, {Kerzendorf}, {Khvalko}, {King}, {Kirkby}, {Kulkarni}, {Kumar},
  {Lee}, {Lenz}, {Littlefair}, {Ma}, {Macleod}, {Mastropietro}, {McCully},
  {Montagnac}, {Morris}, {Mueller}, {Mumford}, {Muna}, {Murphy}, {Nelson},
  {Nguyen}, {Ninan}, {N{\"o}the}, {Ogaz}, {Oh}, {Parejko}, {Parley}, {Pascual},
  {Patil}, {Patil}, {Plunkett}, {Prochaska}, {Rastogi}, {Reddy Janga},
  {Sabater}, {Sakurikar}, {Seifert}, {Sherbert}, {Sherwood-Taylor}, {Shih},
  {Sick}, {Silbiger}, {Singanamalla}, {Singer}, {Sladen}, {Sooley},
  {Sornarajah}, {Streicher}, {Teuben}, {Thomas}, {Tremblay}, {Turner},
  {Terr{\'o}n}, {van Kerkwijk}, {de la Vega}, {Watkins}, {Weaver}, {Whitmore},
  {Woillez}, {Zabalza}, \& {Astropy Contributors}}]{astropy2}
{Astropy Collaboration}, {Price-Whelan}, A.~M., {Sip{\H{o}}cz}, B.~M., {et~al.}
  2018, \aj, 156, 123, \dodoi{10.3847/1538-3881/aabc4f}

\bibitem[{{Balcells} {et~al.}(2003){Balcells}, {Graham},
  {Dom{\'\i}nguez-Palmero}, \& {Peletier}}]{2003ApJ...582L..79B}
{Balcells}, M., {Graham}, A.~W., {Dom{\'\i}nguez-Palmero}, L., \& {Peletier},
  R.~F. 2003, \apjl, 582, L79, \dodoi{10.1086/367783}

\bibitem[{{Barazza} {et~al.}(2003){Barazza}, {Binggeli}, \& {Jerjen}}]{Barazza}
{Barazza}, F.~D., {Binggeli}, B., \& {Jerjen}, H. 2003, \aap, 407, 121,
  \dodoi{10.1051/0004-6361:20030872}

\bibitem[{Barbary(2016)}]{sep}
Barbary, K. 2016, {SEP: Source Extractor as a library},
  \dodoi{10.21105/joss.00058}

\bibitem[{{Begelman} \& {Rees}(1978)}]{begelman}
{Begelman}, M.~C., \& {Rees}, M.~J. 1978, \mnras, 185, 847,
  \dodoi{10.1093/mnras/185.4.847}

\bibitem[{{Bekki} \& {Couch}(2001)}]{bekki01}
{Bekki}, K., \& {Couch}, W.~J. 2001, \apjl, 557, L19, \dodoi{10.1086/323139}

\bibitem[{Bellovary {et~al.}(2018)Bellovary, Cleary, Munshi, Tremmel,
  Christensen, Brooks, \& Quinn}]{bellovary}
Bellovary, J.~M., Cleary, C.~E., Munshi, F., {et~al.} 2018, Monthly Notices of
  the Royal Astronomical Society, 482, 2913, \dodoi{10.1093/mnras/sty2842}

\bibitem[{{Bertin} \& {Arnouts}(1996)}]{bertin}
{Bertin}, E., \& {Arnouts}, S. 1996, \aaps, 117, 393,
  \dodoi{10.1051/aas:1996164}

\bibitem[{{Binggeli} {et~al.}(2000){Binggeli}, {Barazza}, \&
  {Jerjen}}]{binggeli00}
{Binggeli}, B., {Barazza}, F., \& {Jerjen}, H. 2000, \aap, 359, 447

\bibitem[{{Binggeli} {et~al.}(1984){Binggeli}, {Sandage}, \&
  {Tarenghi}}]{binggeli}
{Binggeli}, B., {Sandage}, A., \& {Tarenghi}, M. 1984, \aj, 89, 64,
  \dodoi{10.1086/113484}

\bibitem[{{Blakeslee} {et~al.}(1997){Blakeslee}, {Tonry}, \&
  {Metzger}}]{blakeslee}
{Blakeslee}, J.~P., {Tonry}, J.~L., \& {Metzger}, M.~R. 1997, \aj, 114, 482,
  \dodoi{10.1086/118488}

\bibitem[{{B{\"o}ker} {et~al.}(2002){B{\"o}ker}, {Laine}, {van der Marel},
  {Sarzi}, {Rix}, {Ho}, \& {Shields}}]{2002AJ....123.1389B}
{B{\"o}ker}, T., {Laine}, S., {van der Marel}, R.~P., {et~al.} 2002, \aj, 123,
  1389, \dodoi{10.1086/339025}

\bibitem[{{B{\"o}ker} {et~al.}(2004){B{\"o}ker}, {Sarzi}, {McLaughlin}, {van
  der Marel}, {Rix}, {Ho}, \& {Shields}}]{boker04}
{B{\"o}ker}, T., {Sarzi}, M., {McLaughlin}, D.~E., {et~al.} 2004, \aj, 127,
  105, \dodoi{10.1086/380231}

\bibitem[{{Bothun} \& {Mould}(1988)}]{bothun}
{Bothun}, G.~D., \& {Mould}, J.~R. 1988, \apj, 324, 123, \dodoi{10.1086/165885}

\bibitem[{{Burkert} \& {Forbes}(2020)}]{burkert}
{Burkert}, A., \& {Forbes}, D.~A. 2020, \aj, 159, 56,
  \dodoi{10.3847/1538-3881/ab5b0e}

\bibitem[{{Caldwell}(1983)}]{caldwell}
{Caldwell}, N. 1983, \aj, 88, 804, \dodoi{10.1086/113367}

\bibitem[{{Caldwell} \& {Bothun}(1987)}]{caldwell87}
{Caldwell}, N., \& {Bothun}, G.~D. 1987, \aj, 94, 1126, \dodoi{10.1086/114550}

\bibitem[{{Capuzzo-Dolcetta} \& {Mastrobuono-Battisti}(2009)}]{CDMB2009}
{Capuzzo-Dolcetta}, R., \& {Mastrobuono-Battisti}, A. 2009, \aap, 507, 183,
  \dodoi{10.1051/0004-6361/200912255}

\bibitem[{Carlsten {et~al.}(2022)Carlsten, Greene, Beaton, \&
  Greco}]{Carlsten_2022}
Carlsten, S.~G., Greene, J.~E., Beaton, R.~L., \& Greco, J.~P. 2022, The
  Astrophysical Journal, 927, 44, \dodoi{10.3847/1538-4357/ac457e}

\bibitem[{{C{\^o}t{\'e}} {et~al.}(2006){C{\^o}t{\'e}}, {Piatek}, {Ferrarese},
  {Jord{\'a}n}, {Merritt}, {Peng}, {Ha{\c{s}}egan}, {Blakeslee}, {Mei}, {West},
  {Milosavljevi{\'c}}, \& {Tonry}}]{Cote2006}
{C{\^o}t{\'e}}, P., {Piatek}, S., {Ferrarese}, L., {et~al.} 2006, \apjs, 165,
  57, \dodoi{10.1086/504042}

\bibitem[{{C{\^o}t{\'e}} {et~al.}(2007){C{\^o}t{\'e}}, {Ferrarese},
  {Jord{\'a}n}, {Blakeslee}, {Chen}, {Infante}, {Merritt}, {Mei}, {Peng},
  {Tonry}, {West}, \& {West}}]{cote}
{C{\^o}t{\'e}}, P., {Ferrarese}, L., {Jord{\'a}n}, A., {et~al.} 2007, \apj,
  671, 1456, \dodoi{10.1086/522822}

\bibitem[{{de Vaucouleurs}(1948)}]{dev}
{de Vaucouleurs}, G. 1948, Annales d'Astrophysique, 11, 247

\bibitem[{{den Brok} {et~al.}(2014){den Brok}, {Peletier}, {Seth}, {Balcells},
  {Dominguez}, {Graham}, {Carter}, {Erwin}, {Ferguson}, {Goudfrooij},
  {Guzm{\'a}n}, {Hoyos}, {Jogee}, {Lucey}, {Phillipps}, {Puzia}, {Valentijn},
  {Verdoes Kleijn}, \& {Weinzirl}}]{denBrok2014}
{den Brok}, M., {Peletier}, R.~F., {Seth}, A., {et~al.} 2014, \mnras, 445,
  2385, \dodoi{10.1093/mnras/stu1906}

\bibitem[{{Dey} {et~al.}(2019){Dey}, {Schlegel}, {Lang}, {Blum}, {Burleigh},
  {Fan}, {Findlay}, {Finkbeiner}, {Herrera}, {Juneau}, {Landriau}, {Levi},
  {McGreer}, {Meisner}, {Myers}, {Moustakas}, {Nugent}, {Patej}, {Schlafly},
  {Walker}, {Valdes}, {Weaver}, {Y{\`e}che}, {Zou}, {Zhou}, {Abareshi},
  {Abbott}, {Abolfathi}, {Aguilera}, {Alam}, {Allen}, {Alvarez}, {Annis},
  {Ansarinejad}, {Aubert}, {Beechert}, {Bell}, {BenZvi}, {Beutler}, {Bielby},
  {Bolton}, {Brice{\~n}o}, {Buckley-Geer}, {Butler}, {Calamida}, {Carlberg},
  {Carter}, {Casas}, {Castander}, {Choi}, {Comparat}, {Cukanovaite}, {Delubac},
  {DeVries}, {Dey}, {Dhungana}, {Dickinson}, {Ding}, {Donaldson}, {Duan},
  {Duckworth}, {Eftekharzadeh}, {Eisenstein}, {Etourneau}, {Fagrelius},
  {Farihi}, {Fitzpatrick}, {Font-Ribera}, {Fulmer}, {G{\"a}nsicke},
  {Gaztanaga}, {George}, {Gerdes}, {Gontcho}, {Gorgoni}, {Green}, {Guy},
  {Harmer}, {Hernand ez}, {Honscheid}, {Huang}, {James}, {Jannuzi}, {Jiang},
  {Joyce}, {Karcher}, {Karkar}, {Kehoe}, {Kneib}, {Kueter-Young}, {Lan},
  {Lauer}, {Le Guillou}, {Le Van Suu}, {Lee}, {Lesser}, {Perreault Levasseur},
  {Li}, {Mann}, {Marshall}, {Mart{\'\i}nez-V{\'a}zquez}, {Martini}, {du Mas des
  Bourboux}, {McManus}, {Meier}, {M{\'e}nard}, {Metcalfe},
  {Mu{\~n}oz-Guti{\'e}rrez}, {Najita}, {Napier}, {Narayan}, {Newman}, {Nie},
  {Nord}, {Norman}, {Olsen}, {Paat}, {Palanque-Delabrouille}, {Peng},
  {Poppett}, {Poremba}, {Prakash}, {Rabinowitz}, {Raichoor}, {Rezaie},
  {Robertson}, {Roe}, {Ross}, {Ross}, {Rudnick}, {Safonova}, {Saha},
  {S{\'a}nchez}, {Savary}, {Schweiker}, {Scott}, {Seo}, {Shan}, {Silva},
  {Slepian}, {Soto}, {Sprayberry}, {Staten}, {Stillman}, {Stupak}, {Summers},
  {Sien Tie}, {Tirado}, {Vargas-Maga{\~n}a}, {Vivas}, {Wechsler}, {Williams},
  {Yang}, {Yang}, {Yapici}, {Zaritsky}, {Zenteno}, {Zhang}, {Zhang}, {Zhou}, \&
  {Zhou}}]{dey}
{Dey}, A., {Schlegel}, D.~J., {Lang}, D., {et~al.} 2019, \aj, 157, 168,
  \dodoi{10.3847/1538-3881/ab089d}

\bibitem[{{Eigenthaler} {et~al.}(2018){Eigenthaler}, {Puzia}, {Taylor},
  {Ordenes-Brice{\~n}o}, {Mu{\~n}oz}, {Ribbeck}, {Alamo-Mart{\'\i}nez},
  {Zhang}, {{\'A}ngel}, {Capaccioli}, {C{\^o}t{\'e}}, {Ferrarese}, {Galaz},
  {Grebel}, {Hempel}, {Hilker}, {Lan{\c{c}}on}, {Mieske}, {Miller}, {Paolillo},
  {Powalka}, {Richtler}, {Roediger}, {Rong}, {S{\'a}nchez-Janssen}, \&
  {Spengler}}]{Eigenthaler2018}
{Eigenthaler}, P., {Puzia}, T.~H., {Taylor}, M.~A., {et~al.} 2018, \apj, 855,
  142, \dodoi{10.3847/1538-4357/aaab60}

\bibitem[{{Fahrion} {et~al.}(2021){Fahrion}, {Lyubenova}, {van de Ven},
  {Hilker}, {Leaman}, {Falc{\'o}n-Barroso}, {Bittner}, {Coccato}, {Corsini},
  {Gadotti}, {Iodice}, {McDermid}, {Mart{\'\i}n-Navarro}, {Pinna}, {Poci},
  {Sarzi}, {de Zeeuw}, \& {Zhu}}]{Fahrion_2021}
{Fahrion}, K., {Lyubenova}, M., {van de Ven}, G., {et~al.} 2021, \aap, 650,
  A137, \dodoi{10.1051/0004-6361/202140644}

\bibitem[{{Ferrarese} {et~al.}(2006){Ferrarese}, {C{\^o}t{\'e}}, {Dalla
  Bont{\`a}}, {Peng}, {Merritt}, {Jord{\'a}n}, {Blakeslee}, {Ha{\c{s}}egan},
  {Mei}, {Piatek}, {Tonry}, \& {West}}]{ferr}
{Ferrarese}, L., {C{\^o}t{\'e}}, P., {Dalla Bont{\`a}}, E., {et~al.} 2006,
  \apjl, 644, L21, \dodoi{10.1086/505388}

\bibitem[{Forbes {et~al.}(2020)Forbes, Alabi, Romanowsky, Brodie, \&
  Arimoto}]{2020Forbes_GC}
Forbes, D.~A., Alabi, A., Romanowsky, A.~J., Brodie, J.~P., \& Arimoto, N.
  2020, Monthly Notices of the Royal Astronomical Society, 492, 4874,
  \dodoi{10.1093/mnras/staa180}

\bibitem[{{Foreman-Mackey} {et~al.}(2013{\natexlab{a}}){Foreman-Mackey},
  {Hogg}, {Lang}, \& {Goodman}}]{EMCEE}
{Foreman-Mackey}, D., {Hogg}, D.~W., {Lang}, D., \& {Goodman}, J.
  2013{\natexlab{a}}, \pasp, 125, 306, \dodoi{10.1086/670067}

\bibitem[{{Foreman-Mackey} {et~al.}(2013{\natexlab{b}}){Foreman-Mackey},
  {Hogg}, {Lang}, \& {Goodman}}]{FM2013}
---. 2013{\natexlab{b}}, \pasp, 125, 306, \dodoi{10.1086/670067}

\bibitem[{{Georgiev} \& {B{\"o}ker}(2014)}]{Georgiev2014}
{Georgiev}, I.~Y., \& {B{\"o}ker}, T. 2014, \mnras, 441, 3570,
  \dodoi{10.1093/mnras/stu797}

\bibitem[{{Gnedin} {et~al.}(2014){Gnedin}, {Ostriker}, \& {Tremaine}}]{gnedin}
{Gnedin}, O.~Y., {Ostriker}, J.~P., \& {Tremaine}, S. 2014, \apj, 785, 71,
  \dodoi{10.1088/0004-637X/785/1/71}

\bibitem[{{Hinshaw} {et~al.}(2013){Hinshaw}, {Larson}, {Komatsu}, {Spergel},
  {Bennett}, {Dunkley}, {Nolta}, {Halpern}, {Hill}, \& {Odegard}}]{wmap9}
{Hinshaw}, G., {Larson}, D., {Komatsu}, E., {et~al.} 2013, \apjs, 208, 19,
  \dodoi{10.1088/0067-0049/208/2/19}

\bibitem[{{Hoyer} {et~al.}(2021){Hoyer}, {Neumayer}, {Georgiev}, {Seth}, \&
  {Greene}}]{hoyer}
{Hoyer}, N., {Neumayer}, N., {Georgiev}, I.~Y., {Seth}, A.~C., \& {Greene},
  J.~E. 2021, \mnras, 507, 3246, \dodoi{10.1093/mnras/stab2277}

\bibitem[{{Hunter}(2007)}]{matplotlib}
{Hunter}, J.~D. 2007, Computing in Science and Engineering, 9, 90,
  \dodoi{10.1109/MCSE.2007.55}

\bibitem[{{Kadowaki} {et~al.}(2021){Kadowaki}, {Zaritsky}, {Donnerstein}, {RS},
  {Karunakaran}, \& {Spekkens}}]{kadowaki21}
{Kadowaki}, J., {Zaritsky}, D., {Donnerstein}, R.~L., {et~al.} 2021, \apj, 923,
  257, \dodoi{10.3847/1538-4357/ac2948}

\bibitem[{{Kollmeier} {et~al.}(2017){Kollmeier}, {Zasowski}, {Rix}, {Johns},
  {Anderson}, {Drory}, {Johnson}, {Pogge}, {Bird}, {Blanc}, {Brownstein},
  {Crane}, {De Lee}, {Klaene}, {Kreckel}, {MacDonald}, {Merloni}, {Ness},
  {O'Brien}, {Sanchez-Gallego}, {Sayres}, {Shen}, {Thakar}, {Tkachenko},
  {Aerts}, {Blanton}, {Eisenstein}, {Holtzman}, {Maoz}, {Nandra}, {Rockosi},
  {Weinberg}, {Bovy}, {Casey}, {Chaname}, {Clerc}, {Conroy}, {Eracleous},
  {G{\"a}nsicke}, {Hekker}, {Horne}, {Kauffmann}, {McQuinn}, {Pellegrini},
  {Schinnerer}, {Schlafly}, {Schwope}, {Seibert}, {Teske}, \& {van
  Saders}}]{SDSS_Kollmeier}
{Kollmeier}, J.~A., {Zasowski}, G., {Rix}, H.-W., {et~al.} 2017, arXiv
  e-prints, arXiv:1711.03234, \dodoi{10.48550/arXiv.1711.03234}

\bibitem[{{Kravtsov} \& {Gnedin}(2005)}]{kravtsov}
{Kravtsov}, A.~V., \& {Gnedin}, O.~Y. 2005, \apj, 623, 650,
  \dodoi{10.1086/428636}

\bibitem[{{Kruijssen} {et~al.}(2012){Kruijssen}, {Pelupessy}, {Lamers},
  {Portegies Zwart}, {Bastian}, \& {Icke}}]{kruijssen}
{Kruijssen}, J.~M.~D., {Pelupessy}, F.~I., {Lamers}, H. J.~G.~L.~M., {et~al.}
  2012, \mnras, 421, 1927, \dodoi{10.1111/j.1365-2966.2012.20322.x}

\bibitem[{{Lim} {et~al.}(2018){Lim}, {Peng}, {C{\^o}t{\'e}}, {Sales}, {den
  Brok}, {Blakeslee}, \& {Guhathakurta}}]{Lim2018}
{Lim}, S., {Peng}, E.~W., {C{\^o}t{\'e}}, P., {et~al.} 2018, \apj, 862, 82,
  \dodoi{10.3847/1538-4357/aacb81}

\bibitem[{{Lotz} {et~al.}(2001){Lotz}, {Telford}, {Ferguson}, {Miller},
  {Stiavelli}, \& {Mack}}]{lotz}
{Lotz}, J.~M., {Telford}, R., {Ferguson}, H.~C., {et~al.} 2001, \apj, 552, 572,
  \dodoi{10.1086/320545}

\bibitem[{{McKinney}(2010)}]{pandas}
{McKinney}, W. 2010, Proceedings of the 9th Python in Science Conference, 51

\bibitem[{{Mihos} \& {Hernquist}(1994)}]{mihos94}
{Mihos}, J.~C., \& {Hernquist}, L. 1994, \apjl, 437, L47,
  \dodoi{10.1086/187679}

\bibitem[{{Miller} \& {Hamilton}(2002)}]{miller}
{Miller}, M.~C., \& {Hamilton}, D.~P. 2002, \mnras, 330, 232,
  \dodoi{10.1046/j.1365-8711.2002.05112.x}

\bibitem[{{Miller} \& {Smith}(1992)}]{miller92}
{Miller}, R.~H., \& {Smith}, B.~F. 1992, \apj, 393, 508, \dodoi{10.1086/171523}

\bibitem[{{Millman} \& {Aivazis}(2011)}]{scipy2}
{Millman}, K.~J., \& {Aivazis}, M. 2011, Computing in Science and Engineering,
  13, 9, \dodoi{10.1109/MCSE.2011.36}

\bibitem[{{Modak} {et~al.}(2023){Modak}, {Danieli}, \& {Greene}}]{modak}
{Modak}, S., {Danieli}, S., \& {Greene}, J.~E. 2023, \apj, 950, 178,
  \dodoi{10.3847/1538-4357/accc2d}

\bibitem[{{Neumayer} {et~al.}(2020){Neumayer}, {Seth}, \&
  {B{\"o}ker}}]{neumayer}
{Neumayer}, N., {Seth}, A., \& {B{\"o}ker}, T. 2020, \aapr, 28, 4,
  \dodoi{10.1007/s00159-020-00125-0}

\bibitem[{{Neumayer} {et~al.}(2011){Neumayer}, {Walcher}, {Andersen},
  {S{\'a}nchez}, {B{\"o}ker}, \& {Rix}}]{2011MNRAS.413.1875N}
{Neumayer}, N., {Walcher}, C.~J., {Andersen}, D., {et~al.} 2011, \mnras, 413,
  1875, \dodoi{10.1111/j.1365-2966.2011.18266.x}

\bibitem[{{Oke}(1964)}]{oke1}
{Oke}, J.~B. 1964, \apj, 140, 689, \dodoi{10.1086/147960}

\bibitem[{{Oke} \& {Gunn}(1983)}]{oke2}
{Oke}, J.~B., \& {Gunn}, J.~E. 1983, \apj, 266, 713, \dodoi{10.1086/160817}

\bibitem[{{Oliphant}(2007)}]{scipy1}
{Oliphant}, T.~E. 2007, Computing in Science and Engineering, 9, 10,
  \dodoi{10.1109/MCSE.2007.58}

\bibitem[{{Peng} {et~al.}(2002){Peng}, {Ho}, {Impey}, \& {Rix}}]{peng}
{Peng}, C.~Y., {Ho}, L.~C., {Impey}, C.~D., \& {Rix}, H.-W. 2002, \aj, 124,
  266, \dodoi{10.1086/340952}

\bibitem[{Peng {et~al.}(2010)Peng, Ho, Impey, \& Rix}]{Peng_2010}
Peng, C.~Y., Ho, L.~C., Impey, C.~D., \& Rix, H.-W. 2010, The Astronomical
  Journal, 139, 2097, \dodoi{10.1088/0004-6256/139/6/2097}

\bibitem[{{Poulain} {et~al.}(2021){Poulain}, {Marleau}, {Habas}, {Duc},
  {S{\'a}nchez-Janssen}, {Durrell}, {Paudel}, {Ahad}, {Chougule}, {M{\"u}ller},
  {Lim}, {B{\'\i}lek}, \& {Fensch}}]{Poulain}
{Poulain}, M., {Marleau}, F.~R., {Habas}, R., {et~al.} 2021, \mnras, 506, 5494,
  \dodoi{10.1093/mnras/stab2092}

\bibitem[{{Prasad} \& {Jog}(2017)}]{prasad}
{Prasad}, C., \& {Jog}, C.~J. 2017, \aap, 600, A17,
  \dodoi{10.1051/0004-6361/201630071}

\bibitem[{{Prole} {et~al.}(2018){Prole}, {Davies}, {Keenan}, \&
  {Davies}}]{prole}
{Prole}, D.~J., {Davies}, J.~I., {Keenan}, O.~C., \& {Davies}, L.~J.~M. 2018,
  \mnras, 478, 667, \dodoi{10.1093/mnras/sty1021}

\bibitem[{{Renaud}(2018)}]{renaud}
{Renaud}, F. 2018, \nar, 81, 1, \dodoi{10.1016/j.newar.2018.03.001}

\bibitem[{{Rossa} {et~al.}(2006){Rossa}, {van der Marel}, {B{\"o}ker},
  {Gerssen}, {Ho}, {Rix}, {Shields}, \& {Walcher}}]{rossa}
{Rossa}, J., {van der Marel}, R.~P., {B{\"o}ker}, T., {et~al.} 2006, \aj, 132,
  1074, \dodoi{10.1086/505968}

\bibitem[{{S{\'a}nchez-Janssen} {et~al.}(2019){S{\'a}nchez-Janssen},
  {C{\^o}t{\'e}}, {Ferrarese}, {Peng}, {Roediger}, {Blakeslee}, {Emsellem},
  {Puzia}, {Spengler}, {Taylor}, {{\'A}lamo-Mart{\'\i}nez}, {Boselli},
  {Cantiello}, {Cuillandre}, {Duc}, {Durrell}, {Gwyn}, {MacArthur},
  {Lan{\c{c}}on}, {Lim}, {Liu}, {Mei}, {Miller}, {Mu{\~n}oz}, {Mihos},
  {Paudel}, {Powalka}, \& {Toloba}}]{sanchez}
{S{\'a}nchez-Janssen}, R., {C{\^o}t{\'e}}, P., {Ferrarese}, L., {et~al.} 2019,
  \apj, 878, 18, \dodoi{10.3847/1538-4357/aaf4fd}

\bibitem[{{S{\'a}nchez-Salcedo} \& {Lora}(2022)}]{salcedo}
{S{\'a}nchez-Salcedo}, F.~J., \& {Lora}, V. 2022, \mnras, 511, 1860,
  \dodoi{10.1093/mnras/stac170}

\bibitem[{{Scott} \& {Graham}(2013)}]{2013ApJ...763...76S}
{Scott}, N., \& {Graham}, A.~W. 2013, \apj, 763, 76,
  \dodoi{10.1088/0004-637X/763/2/76}

\bibitem[{Sugiura(1978)}]{sugiura}
Sugiura, N. 1978, Communications in Statistics - Theory and Methods, 7, 13,
  \dodoi{10.1080/03610927808827599}

\bibitem[{Taga \& Iye(1998)}]{taga}
Taga, M., \& Iye, M. 1998, Monthly Notices of the Royal Astronomical Society,
  299, 111, \dodoi{10.1046/j.1365-8711.1998.01753.x}

\bibitem[{{Tremaine}(1976)}]{tremaine}
{Tremaine}, S.~D. 1976, \apj, 203, 345, \dodoi{10.1086/154085}

\bibitem[{{Turner} {et~al.}(2012){Turner}, {C{\^o}t{\'e}}, {Ferrarese},
  {Jord{\'a}n}, {Blakeslee}, {Mei}, {Peng}, \& {West}}]{turner}
{Turner}, M.~L., {C{\^o}t{\'e}}, P., {Ferrarese}, L., {et~al.} 2012, \apjs,
  203, 5, \dodoi{10.1088/0067-0049/203/1/5}

\bibitem[{{van der Walt} {et~al.}(2011){van der Walt}, {Colbert}, \&
  {Varoquaux}}]{numpy}
{van der Walt}, S., {Colbert}, S.~C., \& {Varoquaux}, G. 2011, Computing in
  Science and Engineering, 13, 22, \dodoi{10.1109/MCSE.2011.37}

\bibitem[{van~der Walt {et~al.}(2014)van~der Walt, {S}ch\"onberger,
  {Nunez-Iglesias}, {B}oulogne, {W}arner, {Y}ager, {G}ouillart, {Y}u, \& the
  scikit-image contributors}]{scikit-image}
van~der Walt, S., {S}ch\"onberger, J.~L., {Nunez-Iglesias}, J., {et~al.} 2014,
  PeerJ, 2, e453, \dodoi{10.7717/peerj.453}

\bibitem[{{van Dokkum} {et~al.}(2015){van Dokkum}, {Abraham}, {Merritt},
  {Zhang}, {Geha}, \& {Conroy}}]{2015vanDokkum}
{van Dokkum}, P.~G., {Abraham}, R., {Merritt}, A., {et~al.} 2015, \apjl, 798,
  L45, \dodoi{10.1088/2041-8205/798/2/L45}

\bibitem[{{Wehner} \& {Harris}(2006)}]{wehner}
{Wehner}, E.~H., \& {Harris}, W.~E. 2006, \apjl, 644, L17,
  \dodoi{10.1086/505387}

\bibitem[{{Zaritsky}(2022)}]{2022Zaritsky_GC}
{Zaritsky}, D. 2022, \mnras, 513, 2609, \dodoi{10.1093/mnras/stac1072}

\bibitem[{{Zaritsky} {et~al.}(2023){Zaritsky}, {Donnerstein}, {Dey},
  {Karunakaran}, {Kadowaki}, {Khim}, {Spekkens}, \& {Zhang}}]{smudges5}
{Zaritsky}, D., {Donnerstein}, R., {Dey}, A., {et~al.} 2023, \apjs, 267, 27,
  \dodoi{10.3847/1538-4365/acdd71}

\bibitem[{{Zaritsky} {et~al.}(2021){Zaritsky}, {Donnerstein}, {Karunakaran},
  {Barbosa}, {Dey}, {Kadowaki}, {Spekkens}, \& {Zhang}}]{smudges2}
{Zaritsky}, D., {Donnerstein}, R., {Karunakaran}, A., {et~al.} 2021, \apjs,
  257, 60, \dodoi{10.3847/1538-4365/ac2607}

\bibitem[{{Zaritsky} {et~al.}(2022){Zaritsky}, {Donnerstein}, {Karunakaran},
  {Barbosa}, {Dey}, {Kadowaki}, {Spekkens}, \& {Zhang}}]{smudges3}
---. 2022, \apjs, 261, 11, \dodoi{10.3847/1538-4365/ac6ceb}

\bibitem[{{Zaritsky} {et~al.}(2019){Zaritsky}, {Donnerstein}, {Dey},
  {Kadowaki}, {Zhang}, {Karunakaran}, {Mart{\'\i}nez-Delgado}, {Rahman}, \&
  {Spekkens}}]{smudges}
{Zaritsky}, D., {Donnerstein}, R., {Dey}, A., {et~al.} 2019, \apjs, 240, 1,
  \dodoi{10.3847/1538-4365/aaefe9}

\end{thebibliography}
\bibliographystyle{aasjournal}

\end{document}